\documentclass{aastex}

\usepackage{apjfonts}
\usepackage{emulateapj5}

\shorttitle{Furlanetto \& Loeb}

\submitted{ApJ, Submitted}

\newcommand{\bq}{\begin{equation}}
\newcommand{\eq}{\end{equation}}
\newcommand{\hunits}{\mbox{ km s$^{-1}$ Mpc$^{-1}$}}
\newcommand{\kms}{\mbox{ km s$^{-1}$}}
\newcommand{\fluxunits}{\mbox{ erg s$^{-1}$ cm$^{-2}$ Hz$^{-1}$ sr$^{-1}$}}
\newcommand{\ergs}{\mbox{ erg s$^{-1}$}}
\newcommand{\secinv}{\mbox{ s$^{-1}$}}
\newcommand{\colden}{\mbox{ cm$^{-2}$}}
\newcommand{\erg}{\mbox{ erg}}
\newcommand{\kpc}{\mbox{ kpc}}
\newcommand{\yr}{\mbox{ yr}}
\newcommand{\hr}{\mbox{ hr}}

\newcommand{\kel}{\mbox{ K}}
\newcommand{\ev}{\mbox{ eV}}
\newcommand{\microjy}{\mbox{ $\mu$Jy}}
\newcommand{\angstrom}{\mbox{ \AA}}
\newcommand{\msun}{\mbox{ M$_\sun$}}
\newcommand{\zsun}{\mbox{ Z$_\sun$}}

\begin{document}

\title{Metal Absorption Lines as Probes of the 
  Intergalactic Medium Prior to the Reionization Epoch}

\author{Steven R. Furlanetto\altaffilmark{1} \& Abraham
Loeb\altaffilmark{1,2,3}} 

\email{sfurlanetto@cfa.harvard.edu, loeb@ias.edu}

\altaffiltext{1} {Harvard-Smithsonian Center for Astrophysics, 60
Garden St., Cambridge, MA 02138 }

\altaffiltext{2}{Institute for Advanced Study, Princeton, NJ 08540}

\altaffiltext{3}{Guggenheim Fellow}

\begin{abstract}
  
  Winds from star-forming galaxies provide the most promising
  explanation for the enrichment of the intergalactic medium with
  heavy elements.  Theoretical and observational arguments indicate
  that the pollution may have occurred at $z \ga 6$; however, direct
  observational tests of such a scenario are needed.  We model
  starburst winds in the high-redshift universe and find that the
  fraction of space filled by enriched material varies strongly with
  the assumed star formation efficiency $f_*$ and the fraction of
  supernova energy powering each wind, $f_{\rm esc}$.  We show that
  metals carried by these winds can be seen in absorption against
  bright background sources, such as quasars or gamma-ray bursts, in
  narrow lines with characteristic equivalent widths $\sim 0.5 \la W
  \la 5 \angstrom$.  We argue that a substantial fraction of the
  metals in high-redshift winds are likely to reside in low ionization
  states (\ion{C}{2}, \ion{O}{1}, \ion{Si}{2}, and \ion{Fe}{2}), but
  higher ionization states (\ion{C}{4} and \ion{Si}{4}) could also
  provide useful probes of the winds.  The number of such lines can
  constrain both $f_*$ and $f_{\rm esc}$.  Statistics of metal
  absorption lines can also be used to identify whether H$_2$ is an
  efficient coolant in the early universe and to study the initial
  mass function of stars at high redshifts.

\end{abstract}

\keywords{ cosmology: theory -- intergalactic medium -- galaxies:
  high-redshift } 

\section{Introduction}

It is now obvious that mechanical feedback from galaxies and quasars
has disturbed a significant fraction of the intergalactic medium
(IGM).  Types of feedback include relativistic quasar jets, winds from
quasars, and winds from star-forming galaxies.  These mechanisms have
had a variety of consequences, including IGM heating (e.g.,
\citealt{voit}), the suppression of star formation in the host galaxy
(e.g., \citealt{springel02}) and in nearby galaxies (e.g.,
\citealt{scann01}), and the generation of intergalactic magnetic
fields \citep{kronberg,furl01}.  Perhaps most importantly, feedback is
responsible for enriching the IGM with heavy elements.  Observations
of the Ly$\alpha$ forest at $z \sim 3$ show that absorption systems
with column densities $N_{\rm HI} \ga 10^{14.5} \colden$
(corresponding to overdensities $\delta \ga 5$; \citealt{schaye}) have
metallicities $Z \sim 10^{-3}$--$10^{-2} \zsun$
\citep{tytler,cowie95,songaila96,ellison00}.  Absorbers with $N_{\rm
HI} \la 10^{14.5} \colden$ require stacking or pixel-by-pixel analysis
because of the weakness of the corresponding metal features, but
recent studies suggest similar metallicities down to systems near the
mean density of the IGM
\citep{cowie98,ellison99,ellison00,aguirre-pixel}.  These observations
require that $\ga 5\%$ of space be enriched with heavy elements by $z
\sim 3$ \citep{madau}.  The era at which this pollution occurred is
uncertain.  \citet{songaila} found no evidence for evolution in the
total mass-weighted abundance of \ion{C}{4} and \ion{Si}{4} between $z
\sim 1.5$--$5$ (see also \citealt{qsw}).  However, such mass-weighted
measurements constrain only the metals in the most massive systems and
tell us little about the total volume of enriched space.  Furthermore,
conclusions about the total metallicity based on single ionization
states are fraught with uncertainty.

The history of metal enrichment has several important consequences for
structure formation.  Most directly, its extent measures the volume
over which galaxies have influenced their surroundings
hydrodynamically, which in turn defines their feedback on the
Ly$\alpha$ forest.  Additionally, an early phase of metal injection
may qualitatively change the characteristic mass of star formation
\citep{bromm01,schneider,mackey}.  The transition between
zero-metallicity star formation and ``normal'' star formation has
important implications, e.g., for the expected redshift distribution
of gamma-ray bursts (GRBs; \citealt{bromm-grb}), for reionization
\citep{wyithe,cen}, and for the chemical abundance patterns of
low-metallicity stars \citep{qian01,qian02}.

Stars produced metals inside galaxies, and one must invoke some type
of outflow to transport them throughout the IGM.  Simple dynamical
removal cannot pollute the large regions we observe
(\citealt{aguirre-full}; but see \citealt{gnedin}).  Radiation
pressure-driven dust outflows may suffice \citep{aguirre-dust},
although their effectiveness depends on many untested assumptions
about the outflows, such as the geometry of the magnetic field.
Quasar winds are unlikely by themselves to entrain enough metals to
account for the observed metallicities, although they can help to
disperse the metals over large volumes.  Winds from star-forming
galaxies are therefore the most likely mechanism.  Correlated
supernovae in starbursts can power ``superwinds'' with velocities
$\sim 100$--$1000 \kms$ that reach far outside the host galaxy.  Such
winds have been studied in many local starburst galaxies
\citep{lehnert,martin} as well as in Lyman-break galaxies (LBGs) at $z
\sim 3$ \citep{pettini01,pettini,shapley}.  Because their velocities
exceed the escape speed from many of the hosts, we expect a subset of
these outflows to penetrate into the general IGM.  Escape is most
likely from dwarf galaxies, because the amount of energy available to
power a wind is proportional to the halo mass $M_h$ while the
gravitational potential confining the wind increases with $M_h^2$.
Thus, starburst winds can naturally account for a high-redshift
episode of metal enrichment, when the characteristic mass of galaxies
is small.

A great deal of theoretical work has gone into studying the galactic
wind pollution mechanism.  Ideally, the enrichment process should be
modeled with cosmological simulations.  To this end, several numerical
simulations have incorporated galactic winds (e.g.,
\citealt{scann01,springel02,theuns}).  Unfortunately, we do not yet
fully understand the processes that generate winds even in nearby,
well-observed starburst galaxies, although simulations are improving
\citep{maclow,mori}.  We must therefore insert simple
parameterizations of the wind into the simulation.  In order to
understand the significance of this choice, most work to date has
proceeded using analytic or semi-analytic methods
\citep{nath,ferrara00,madau,scann}.  Such models have the advantage of
allowing much more flexible parameter studies.  Because the relevant
parameters are poorly constrained at present, the IGM filling factor
(i.e., the fraction of space filled by enriched material) estimated by
these studies varies widely (between $\sim 10^{-4}$ and $\sim 0.5$).
\citet{aguirre-full} made a useful compromise between these two
approaches by inserting winds into already completed cosmological
simulations (see also \citealt{aguirre-dust,aguirre-wind}).  This
allowed a broader parameter study than complete simulations and
retained many of their advantages (such as the clustering of sources),
while sacrificing a detailed treatment of the dynamical effects of the
winds.

The crucial unknown parameters in galactic wind models fall into three
broad categories.  First, the characteristic redshifts and masses of
the wind hosts are critical for setting up simulations with finite
resolution and box size.  Second, the unknown star formation
efficiency at high redshifts has dramatic effects on both the sizes of
individual wind bubbles (because it determines the energy input into
the winds) and on the total amount of metals produced.  Finally, a
variety of questions about the structure of the winds remain
unsettled.  We do not know what fraction of the supernova energy is
lost to radiation before entering the wind, what fraction of the
galaxy mass is entrained by the wind, or the geometry of the winds.
Studies of nearby starbursts can help to answer these questions;
however, winds at high redshifts may have dramatically different
characteristics.  For example, local superwinds are bipolar, with the
wind flowing perpendicular to the galaxy disk.  If star formation
precedes disk formation in the halo, we would expect high-redshift
outflows to be more isotropic \citep{madau}.

Ultimately, we can only answer these questions through observations of
the pollution mechanism.  One promising technique is to study
absorption lines caused by metals in the winds.  In both local
starbursts \citep{phillips,lequeux,heckman,hlsa} and LBGs
\citep{pettini01,pettini,shapley} absorption features can be observed
through moderate resolution spectroscopy of the host galaxy.
Unfortunately, the hosts of high-redshift ($z \ga 6$) starbursts will
be too faint to serve as useful background lights.  We therefore must
use other bright background sources, such as quasars and GRBs, against
which absorption can be seen.  Such background sources have the
distinct advantage of allowing us to probe ``relic'' winds whose host
galaxies no longer actively form stars.  Relic wind bubbles are older
and larger than the young winds in starburst galaxies; they are also
more common than active winds because only a fraction of galaxies host
starbursts at any time.  Thus relic bubbles fill more space than
active winds, and a careful study of their properties is crucial.
Similar observations have been attempted for local starbursts
\citep{norman} and an extension of this technique has been used to
study the effects of LBGs on the IGM \citep{adelberger}.  \citet{oh}
has also proposed using metal absorption lines to study the
reionization process, provided that metals have been injected
relatively uniformly in dense regions by $z \sim 6$.

The neutral fraction of IGM hydrogen is relatively large
($\ga 1\%$) before the completion of reionization at $z_r \ga 6$
\citep{becker,fan}.  The resulting absorption blueward of the
Ly$\alpha$ transition \citep{gunn} will render unobservable any metal
lines with rest-frame wavelength $\lambda_m$ in the range $(1+z_r)/(1+z) <
\lambda_m/\lambda_{\rm Ly \alpha} < 1$, where $\lambda_{\rm Ly
  \alpha}= 1216 \angstrom$ and $z$ is the redshift of the absorbing
system.  A transition below the left-hand limit will be observable
provided that it can be separated from the Ly$\alpha$ forest.
A transition with $\lambda_m > \lambda_{\rm Ly \alpha}$ is
unaffected by neutral hydrogen inside a finite redshift interval
$z_{\rm min} < z < z_s$, where $z_s$ is the redshift of the background
source and
\bq 
1+z_{\rm min} = (1+z_s) \,
\frac{\lambda_{\rm Ly \alpha}}{\lambda_m}.
\label{eq:zwindow}
\eq 
Within this range, metal absorption lines will remain visible.  A
collection of sources with a range of $z_s$ should therefore allow us
to probe winds throughout the high-$z$ IGM.

In this paper, we use a simple semi-analytic model for winds from
star-forming galaxies to study how observations of metal absorption by
wind bubbles in the era before reionization can constrain the metal
enrichment process.  We compute the halo star formation histories and
follow the expansion of their winds into the IGM.  We show that metals
carried by the winds produce observable absorption features when seen
against bright background sources, and that with current and future
instruments, observations of these lines can constrain the star
formation efficiency in the early universe and can distinguish between
different halo cooling scenarios and stellar initial mass functions.

We discuss our supernova wind model in \S 2 and metals in the
high-redshift IGM in \S 3.  We present our predictions for the
statistics of metal absorbers in \S 4 and discuss our results in \S 5.
We assume the standard $\Lambda$CDM cosmological parameters of
$\Omega_m=0.3$, $\Omega_\Lambda=0.7$, $\Omega_b=0.05$, $\sigma_8=0.9$,
$n=1$, and a Hubble constant $H_0=100 h \hunits$.  We set $h=0.7$ in
our numerical calculations.

\section{Supernova Wind Model}

In this section we describe our model for superwinds in detail.  We
discuss the abundance of wind sources and their star formation
histories in \S 2.1.  We fix the mechanical power driving the winds in
\S 2.2, and we describe our model for their expansion into the IGM in
\S 2.3.

\subsection{Halo Abundances and Star Formation Histories}

At any redshift, the halo mass function $dn/dM$ determines the number
of wind sources: each star-forming halo has the potential to drive
such a wind.  We use the \citet{press} formalism to calculate the mass
function, with the modifications suggested by \citet{sheth} to better
match numerical simulations \citep{jenkins}.

A halo can only form stars if its mass exceeds a threshold 
\bq
M_{\rm min}(z) = \max[M_{\rm fil}(z),M_{\rm cool}(z)].
\label{eq:mmin}
\eq The first condition requires that the halo mass exceeds the
``filter mass'' $M_{\rm fil}$, or the effective Jeans mass in the
evolving IGM \citep{gnedin-filter}.  The second condition requires
halo gas to cool efficiently after virialization.  In the low
metallicity halos we study, line transitions of atomic and molecular
hydrogen are the only effective coolants.  Atomic cooling is efficient
in halos with $T_{\rm vir} \ga 10^4 \kel$, where the virial
temperature $T_{\rm vir}$ is defined in equation (26) of
\citet{barkana}.  In most of our work, we will fix $M_{\rm cool}$ by
requiring that atomic cooling can proceed.  The vibrational lines of
H$_2$ make it an efficient coolant in halos with $T_{\rm vir} \ga 400
\kel$.  It is currently unknown when (and even whether) H$_2$ is
destroyed by photons below the ionization threshold of atomic hydrogen
\citep{haiman-h2,haiman-abel}, or whether radiative feedback from
early generations of stars enhances the formation of H$_2$
\citep{ricotti}.  We therefore also show results in which H$_2$
cooling is permitted.

The power driving a wind is ultimately determined by the star
formation history of the host halo, which is usually associated with the
halo merger history.  We use a Monte Carlo implementation of the
extended Press-Schechter formalism \citep{lacey} to choose histories
for each halo.  Given a halo of mass $M_h$ at redshift $z_f$, this
algorithm determines the probability distribution of progenitor masses
($M_1$, $M_h-M_1$) that merged to form the object in a small redshift
interval $dz$ around $z_f$ (we choose $dz$ such that the probability
of a major merger is $< 1\%$).  For each halo, we construct the merger
history by first choosing $M_1$ from this distribution.  We then
choose the larger of the two progenitor masses (i.e., $M_1$ if $M_1 >
M_h/2$ and $M_h-M_1$ otherwise) and repeat the process about a redshift
$z_f+dz$.  We continue this procedure until the progenitor mass is
smaller than $M_{\rm min}$ at a redshift $z_{\rm form}$.

In order to simplify our calculations, we divide the merger history
into discrete events in which the halo mass has grown by $> 50\%$.
(We find that our results are insensitive to the value of this
threshold.)  We assume that in each such event a fraction $f_*$ of the
accreted baryonic mass $M_g$ is transformed into stars on a timescale
$t_{\rm sf}$.  We assume that each merger occurs over the free-fall
time of the host halo $t_{\rm ff}$, and we normally set $t_{\rm sf} =
t_{\rm ff}$.  In the rare occasions in which another merger event
begins within $t_{\rm ff}$ of the first, we set $t_{\rm sf}$ equal to
the time between these events.  In fact, our results are relatively
insensitive to $t_{\rm sf}$.  Although the wind dynamics do change
during the ``active'' phase of the wind, most of the expansion occurs
after active star formation has ended as long as $t_{\rm sf} \ll
H(z)^{-1}$.  If this condition is satisfied, the resulting bubble
sizes do not depend strongly on $t_{\rm sf}$ because bubbles spend
most of their time in the late ``remnant'' stages.  However, if
$t_{\rm sf} \sim H(z)^{-1}$, a significant amount of energy is lost to
radiative cooling.  In this case the total volume filled by
metal-enriched winds can decrease by $\sim 25\%$.

With this prescription, we implicitly assume that a progenitor does not
form stars and release mechanical energy until it joins the ``trunk''
of the merger tree.  In reality, as long as the progenitor mass
exceeds $M_{\rm min}(z)$, it would have formed stars before merging
and thus be surrounded by its own wind bubble.  The two bubbles will
presumably merge along with the galaxies in some complicated manner.
Our prescription is reasonable so long as the total wind energy is
approximately conserved during the merging process.  Note also that by
choosing to follow only one of the progenitors at each branching, we
do not necessarily track the largest single progenitor existing at
each step.  This error is not important so long as our Monte Carlo
prescription includes a sufficiently large number of trials.

\subsection{Wind Luminosity}

Once we have fixed the rate of star formation $f_* M_g/t_{\rm sf}$,
the mechanical luminosity in the wind depends only on the initial mass
function (IMF) of the stars and the fraction of supernova energy
available to power the wind.  The IMF dependence can be characterized
entirely by the total amount of stellar mass that must be formed in
order to produce one supernova, $\omega_{\rm SN}$.  For a
\citet{scalo} IMF with lower and upper mass cutoffs $M_l=0.1 \msun$
and $M_u=100 \msun$, $\omega_{\rm SN}=126 \msun$, while for a
\citet{salpeter} IMF with the same mass cutoffs, $\omega_{\rm SN}=135
\msun$.  We use a Scalo IMF in the following because it better matches
observations of the IMF in the local universe; however, the results
are clearly insensitive to this choice.

Recent simulations of the formation of metal-free stars indicate that
these objects may be very different from nearby stars, with typical
masses $\sim 10^2$--$10^3 \msun$ \citep{abel,bromm02}.  This mass
scale is determined by the physics of H$_2$ cooling; we therefore
assume that when molecular hydrogen is an active coolant, star
formation proceeds with a \emph{Very Massive Star (VMS) IMF}.
Unfortunately, these simulations do not yet have the dynamic range to
follow the collapse to nuclear burning, so they are unable to predict
the final IMF or the allowed mass range of metal-free stars.  We
therefore assume (somewhat arbitrarily) that the IMF is a
Salpeter-like power law with an index $\beta=2.35$ and mass cutoffs
$M_l=100 \msun$ and $M_u=500 \msun$.  Interestingly, massive,
low-metallicity stars leave very different remnants from ``normal''
stars \citep{heger}.  For a stellar mass less than $130 \msun$ or
greater than $260 \msun$, the star collapses to a black hole without
significant matter or energy ejection.  However, stars with masses
$130 \msun \la M \la 260 \msun$ are subject to the $e^+e^-$ pair
instability and explode without leaving a remnant black hole.  Unlike
normal supernovae, such explosions expel the highly-enriched core
along with the envelope.  This mechanism may thus have significant
consequences for metal enrichment (see \S 3.1 below).  For our VMS
IMF, we find $\omega_{\rm SN}=462 \msun$.

We assume that each supernova releases an energy $E_{\rm SN}=
10^{51}E_{51} \erg$ into the interstellar medium of the host galaxy.
For a Scalo IMF, we assume $E_{51}=1$.  For a VMS IMF, we assume
$E_{51}=10$, an approximate logarithmic average of the mass-dependent
explosion energy in the models of \citet{heger}.  However, only a
fraction $f_{\rm esc}$ of this energy is available to power the wind:
the rest is lost, primarily to radiative cooling, in the early stages
of the outflow.  Observations of the cool, dense ejecta of starbursts
in the local universe require $f_{\rm esc} \ga 0.1$ \citep{hlsa},
while observations of the warm and hot plasma in the wind imply
$f_{\rm esc} \ga 0.3$ \citep{strickland}.  However, at high redshifts
the characteristic density is much higher, and radiative cooling could
in principle be more important.  Simulations of high-redshift
starbursts suggest that $f_{\rm esc} \sim 0.25$ \citep{mori}.

With these parameters, the mechanical luminosity driving the wind is
\bq L_{w} = 10^{39} \left(
  \frac{f_*}{0.1} \frac{f_{\rm esc}}{0.25} E_{51} \frac{126
    \msun}{\omega_{\rm SN}}
  \frac{10^8 \yr}{t_{\rm sf}} \frac{M_g}{10^7 \msun} \right) \ergs.
\label{eq:lsn}
\eq

\subsection{Wind Model}

We can now calculate the expansion of the wind into the IGM.  The wind
begins when the progenitor mass first exceeds $M_{\rm min}(z)$, and
energy is added to it during each merger event.  We
use the thin-shell approximation \citep{tegmark, furl01}, which is
well-justified in a cosmological context \citep{ikeuchi}.  The
expanding wind sweeps a fraction $1-f_m$ of the ambient IGM into a
thin shell.  The remaining gas leaks into the hot, rarefied interior
whose pressure drives the shell expansion.  Assuming spherical
symmetry, we can then describe the outflow through the following
system of equations:
\begin{eqnarray}
\ddot{R} & = & \frac{4 \pi R^2}{M_s} (p-p_{\rm ext}) - \frac{G}{R^2}
  \left( M_d + M_{\rm gal} + \frac{M_s}{2} \right) 
\nonumber \\
& & + \Omega_\Lambda(z)
  H^2(z) R - \frac{\dot{M}_s}{M_s} \left( \dot{R} - v_{\rm inf} \right),
\label{eq:shellacc}
\\
\dot{p} & = & \frac{L}{2 \pi R^3} - 5 p \frac{\dot{R}}{R},
\label{eq:shellpress}
\\
\dot{M}_s & = &
\left\{
\begin{array}{ll}
0, & v_{\rm inf} \geq \dot{R}, \\
4 \pi R^2 \rho_g (\dot{R} - v_{\rm inf}), & v_{\rm inf} < \dot{R}. 
\end{array}
\right.
\label{eq:shellmass}
\end{eqnarray}
Here, $R$ is the shell radius in physical units, $M_s$ is the shell
mass, and $p$ is the pressure of the hot bubble interior.  The shell
expands through the host halo and into the IGM; the ambient baryon
density, infall velocity field, and external pressure are denoted
$\rho_g$, $v_{\rm inf}$, and $p_{\rm ext}$, respectively.  We assume
that the host galaxy has photoionized the external medium, which is
likely to be true during most of the wind's expansion.  The enclosed
dark matter mass is $M_d$ and the host galaxy mass (which can be time
dependent as the wind entrains material from the host) is $M_{\rm
gal}$.  We use the universal halo profile of \citet{nfw} inside the
accretion shock radius and the self-similar solution of \citet{bert}
outside of this radius, although the details of the profile are not
critical to our results.  Each halo is therefore embedded in a
relatively small overdense infall region, which is in turn surrounded
by IGM at the mean cosmic density.  Both $M_h$ and the halo profile
evolve as the host galaxy grows through accretion.  Finally, the rate
of energy input $L$ into the bubble interior is \bq L = L_w
\Theta(t-t_{\rm sf}) + L_{\rm diss} - L_{\rm comp},
\label{eq:shelllum}
\eq 
where $L_{\rm comp}$ is the inverse Compton cooling luminosity, $L_w$ is the
wind luminosity defined in equation (\ref{eq:lsn}), $\Theta$ is the
Heaviside step function, and
\bq
L_{\rm diss} = - \frac{1}{2} f_d \dot{M} (\dot{R}-v_{\rm inf})^2.
\label{eq:ldiss}
\eq
Here $f_d$ is the fraction of the energy dissipated by the
acceleration of ambient material to the shell velocity that is
retained by the outflow (rather than lost through cooling inside the
shell).  
In the hot ($T \sim 10^6 \kel$), rarefied bubble cavity,
inverse Compton cooling dominates at the high redshifts we consider,
with a cooling time
\bq
\frac{t_{\rm comp}}{H(z)^{-1}} \approx 0.4 \left( \frac{10}{1+z}
\right)^{5/2}.
\label{eq:comptime}
\eq We therefore neglect metal line and free-free cooling inside the
bubble.\footnote{ Here we assume that the ejected metals are
distributed in the same way as the shell and bubble gas, so that only
a fraction $f_m=0.1$ of the metals reside in the hot interior.  Metal
line cooling can then be safely ignored.  If instead \emph{all} the
metals are confined to the hot cavity, metal line cooling cannot be
ignored at $z \la 7$ or when bubbles have cooled in the late stages of
expansion ($T \la 10^5 \kel$); in other situations, inverse Compton
cooling still dominates.  At low redshifts, we note also that heating
by the ultraviolet background radiation field is significant. }
Material in the shell will cool rapidly through either inverse Compton
cooling (when it is ionized) or line cooling (once recombination
occurs).  We therefore set $f_d=0$, although we also show some results
for $f_d=1$.  We describe the expansion model in more detail in
\citet{furl01}.

As emphasized above, the structure of high-redshift winds is not
well-constrained.  They are likely to be more isotropic than those of
local starbursts because their disks are not as well-defined.  LBGs
appear to host isotropic outflows, because wind absorption lines are
seen in \emph{all} LBGs that have been observed \citep{shapley}.
Moreover, the outflow will likely become more spherical as it plows
into the IGM.  We therefore assume that all winds expand spherically.
A second key question is whether star formation occurs before or after
the halo gas collapses into a dense medium.  In the former case, the
wind expands into a spherical, relatively low density medium (see,
e.g., \citealt{madau}).  In this ``no disk formation'' model, we
assume that supernovae inject energy into the center of the halo and
that all of the halo gas is swept up by the shell.  Simulations show
that a substantial fraction ($\sim 0.5$) of the gas in the halo may
remain inside the protogalaxy \citep{mori}; however, this choice makes
only a small difference to our results because most of the expansion
occurs after the swept-up IGM mass exceeds the halo mass (see \S 4.3).

In most of our work, we assume that the halo gas collapses to a dense
medium before the starburst begins, leaving the rest of the halo empty
of baryons.  We assume that the dense medium has a scale length 
\bq
R_d= \frac{\lambda}{\sqrt{2}} R_{\rm vir},
\label{eq:diskscale}
\eq where $R_{\rm vir}$ is the virial radius of the halo as defined in
equation (24) of \citet{barkana} and $\lambda=0.05$ is the spin
parameter \citep{mo}.  With a slight abuse of terminology, we refer to
this case as the ``disk formation'' scenario.  Note that we do not
require well-defined disks to form, only that the gas has collapsed to
a dense central object.  We assume that the wind entrains a fraction
$f_{\rm sw}$ of the protogalactic material.  Motivated by observations
of local disk starbursts, we set $f_{\rm sw}=2 f_*$ in most of what
follows \citep{martin}.\footnote{Note, however, that the inferred star
formation rates in some high-$z$ galaxies (e.g., \citealt{pettini01})
are much larger than those of low-$z$ starbursts.  The calibration of
\citet{martin} may therefore not apply to these systems.  Fortunately,
our results are relatively insensitive to this choice (see Figure
\ref{fig:ewparams}).}  The initial velocity of the outflow $v_w$ is
then fixed (c.f., \citealt{aguirre-full,springel02}): \bq v_w = 310
\left( \frac{f_{\rm esc}}{0.25} \, \frac{2}{f_{\rm sw}/f_*} \,
\frac{126 \msun}{\omega_{\rm SN}} \right)^{1/2} \kms
\label{eq:vdisk}
\eq 
Note that $v_w$ is independent of halo mass, implying that winds will
most easily escape the gravitational attraction of smaller halos.
The wind velocities in nearby starbursts are inferred to be
independent of the host mass \citep{martin}.  

Our prescription for the initial conditions depends on whether disk
formation is allowed.  If it is not, we use the self-similar wind-fed
solution of \citet{ostriker} to calculate the time for the wind to
travel to $R_d$ assuming a constant density within this region.
(Although $R_d$ has no physical meaning in this case, we use it to fix
numbers.)  The self-similar solution also yields the initial velocity,
and we assume that the remaining supernova energy goes into the bubble
pressure.  While this prescription makes several simplifying
assumptions, the results are insensitive to the precise initial
conditions in this case.  If disk formation is allowed, we assume that
the wind travels with a constant velocity $v_w$ within the dense
collapsed region.  The initial time is determined by the time taken
for the wind to reach $R_d$.  For numerical stability, we convert half
of the kinetic energy to the thermal energy of the bubble interior
before beginning the integration.  (The final result is not sensitive
to this conversion.)

When gravity, density variations, external pressure, radiative
cooling, and the cosmological expansion are ignored, our model for a
single explosion recovers the usual Sedov-Taylor solution if $f_d=1$
or the pressure-driven snowplow solution if $f_d=0$ \citep{ostriker}.
With these effects included, we find instead $R \propto E^{0.35}$ at
fixed halo mass, a somewhat steeper energy dependence than expected in
self-similar solutions.  This is principally due to the deceleration
of the Hubble flow: an explosion with smaller energy takes longer to
escape the host halo and therefore must expend more energy
accelerating the swept-up IGM.  Our model matches well with previous
analytic work on outflows (e.g., \citealt{barkana}).  However, it
predicts wind radii approximately twice those found by \citet{madau},
even when we include only the effects that are part of their model.
We do not consider this a cause for concern, because the simulations
of \citet{mori} have a similar discrepancy and compare favorably with
our results.

Figures \ref{fig:radcomp} and \ref{fig:velcomp} show the physical
sizes $R$ and expansion velocities $v_{\rm exp}$ of supernova wind
bubbles at $z=20,12,8,$ and $3$ as a function of host halo mass.  We
assume the following parameters: $f_*=0.1$, $f_{\rm esc} = 0.25$,
$\omega_{\rm SN}=126 \msun$ (corresponding to a Scalo IMF), atomic
cooling, disk formation, and $f_d=0$.  The lower mass limit in each
panel is simply $M_{\rm min}$, and the upper limit is the approximate
mass $M_{\rm max}$ above which the winds cannot escape the host halo.
At a given mass, the dispersion in radius and velocity is caused by
the randomly chosen star formation histories.  For halos near $M_{\rm
min}$, $R$ varies significantly because such halos have undergone only
a few star formation episodes.  Thus, the time elapsed since the wind
began fluctuates strongly.  As $M_h$ increases, the number of mergers
increases and the dispersion decreases.  Recall that the wind begins
when the progenitor crosses the cooling threshold, and each subsequent
merger adds energy to it.  Because massive halos have undergone many
such random merger events, the fluctuations in merger times average
out and the scatter in $R$ is fairly small.  The dispersion increases
again near $M_{\rm max}$, where the star formation history, detailed
wind model, and initial conditions all have a substantial effect on
whether the wind can escape beyond $R_{\rm vir}$.  Fortunately, halos
near $M_{\rm max}$ are sufficiently rare that they have negligible
effects on our final results.

$R$ increases rapidly as redshift decreases because of the increasing
cosmic time available to the outflow.  The dispersion also increases
with redshift, because small variations in the formation time make
larger differences when the total lifetime of the outflow is small.
The weak dependence of radius on mass is primarily due to the
gravitational binding of the host.  Because the energy available to
the starburst scales with $M_h$ while the gravitational potential
scales with $M_h^2$, we expect $R(M_h)$ to flatten as mass increases
until eventually the winds can no longer escape outside the virial
radii of the hosts (shown by the solid curves).  For most of the
redshift range of interest and for reasonable choices of parameters,
the threshold halo mass is $M_{\rm max} \sim 10^{10}$--$3 \times
10^{11} \msun$, increasing as $z$ decreases because halos become less
concentrated.  Well below this threshold mass, we find $R \propto
M_h^{\sim 1/5}$, as would have been expected in a naive Sedov-Taylor
solution (though the agreement is coincidental and the scaling only
approximate).  The top left panel of Figure \ref{fig:radcomp} shows
that LBGs can directly affect the surrounding IGM to (physical)
distances $\sim 100 \kpc$, provided that they reside in halos with
$M_h \sim 10^{11} \msun$
(see also \citealt{aguirre-wind}).  Interestingly, \citet{adelberger}
find that regions within $\sim 125 \kpc$ of LBGs have unexpectedly
small amounts of \ion{H}{1}.  One possible explanation is that winds have
cleared this region of \ion{H}{1}; if so, these observations suggest
that our model provides a reasonable description of the extent of 
outflows.

\centerline{{\vbox{\epsfxsize=8cm\epsfbox{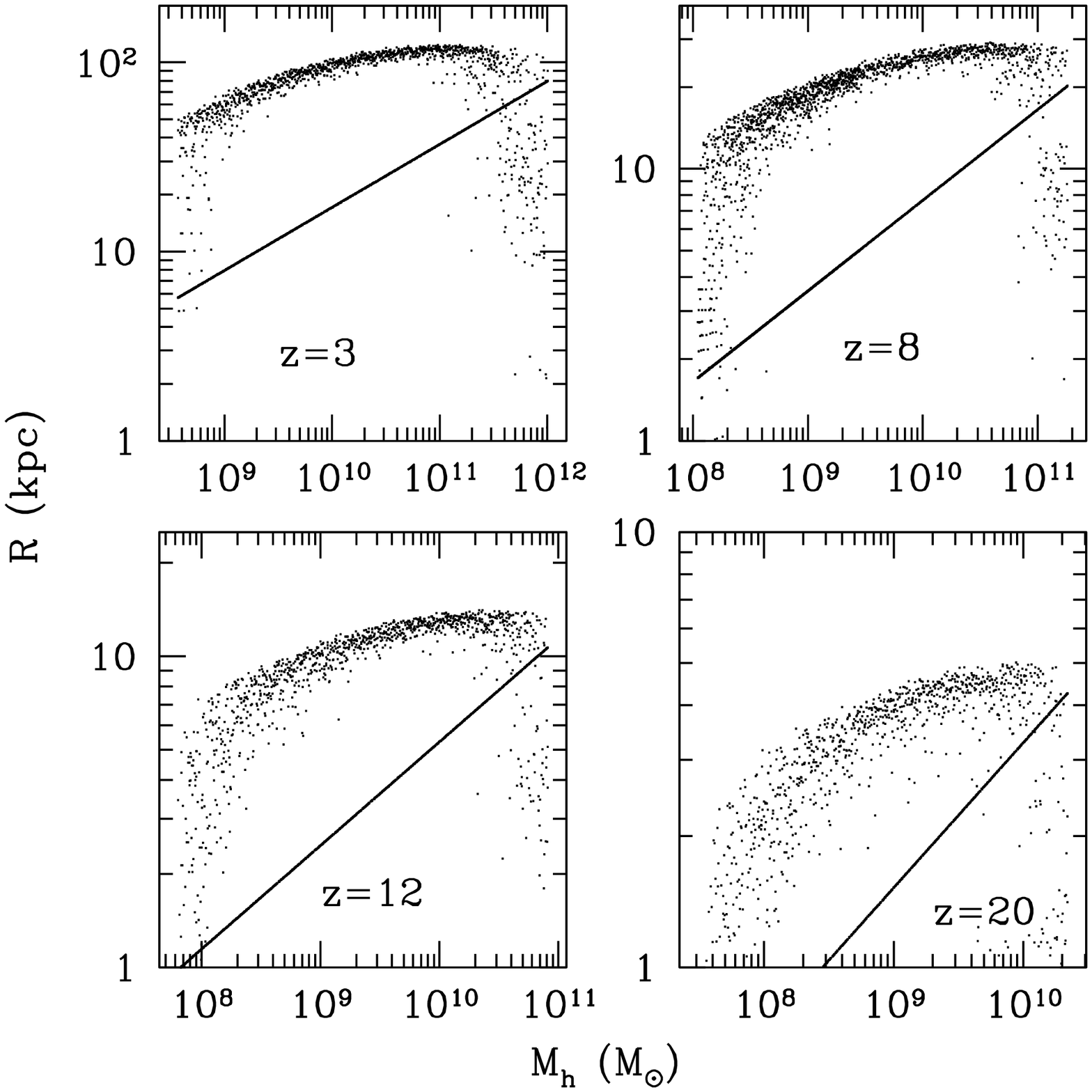}}}}
\figcaption{ Physical radius $R$ of wind bubbles at various redshifts,
as a function of halo mass.  The solid line in each panel shows the
virial radius $R_{\rm vir}$ of the host halo. Note that both the mass
and radius scales change between the different panels.  All panels
assume our standard set of parameters.
\label{fig:radcomp}}
\vskip 0.2in

Figure \ref{fig:velcomp} shows that all of our wind bubbles have
characteristic expansion velocities $v_{\rm exp} \sim 50 \kms$, with a
substantial dispersion.  The velocity increases slowly with mass,
particularly at low redshifts, until $M_h \sim M_{\rm max}$.  Our
expansion velocities are much smaller than those inferred from
observed starburst winds \citep{hlsa,pettini}.  The difference is,
however, easily understood.  While existing observations probe the
early stages of starburst winds, our models primarily sample the late
time behavior.  At early times, the material ejected from the host
galaxy dominates the dynamics, but gravity and collisions with the
ambient IGM inevitably decelerate the winds later on.  We can estimate
$v_{\rm exp}$ by assuming that the host experiences one major merger
per Hubble time and that all of the energy released goes into
accelerating the swept-up medium.  The average expansion velocity is
then
\begin{eqnarray}
v_{\rm exp} & \sim 50 & \kms \left( \frac{f_*}{0.1} \, \frac{f_{\rm
      esc}}{0.25} \, \frac{126 \msun}{\omega_{\rm SN}} \,
  \frac{M_h}{10^9 \msun} \right)^{1/3} 
\nonumber \\
& & \times \left( \frac{20 \kpc}{R}
\right)^{2/3} \left( \frac{10}{1+z} \right)^{1/2}.
\label{eq:vexpavg}
\end{eqnarray} 
This simple estimate neglects cooling, gravity, and the Hubble
flow energy contained in the IGM, but it illustrates the approximate
velocities we find in our model.  Recalling $R \propto M_h^{1/5}$ for
$M_h \ll M_{\rm max}$, equation (\ref{eq:vexpavg}) also predicts a
weak mass dependence $v_{\rm exp} \propto M_h^{1/5}$, near what we
observe in the complete model at small redshifts.  Note that the Mach
number ${\cal M} \equiv v_{\rm exp}/c_s \sim 4$, where $c_s$ is the
sound speed of the photo-ionized IGM.  The thin shell model assumes
${\cal M} \gg 1$, so it is only marginally valid in this regime.  The
detailed expansion may therefore differ somewhat from our model.

\centerline{{\vbox{\epsfxsize=8cm\epsfbox{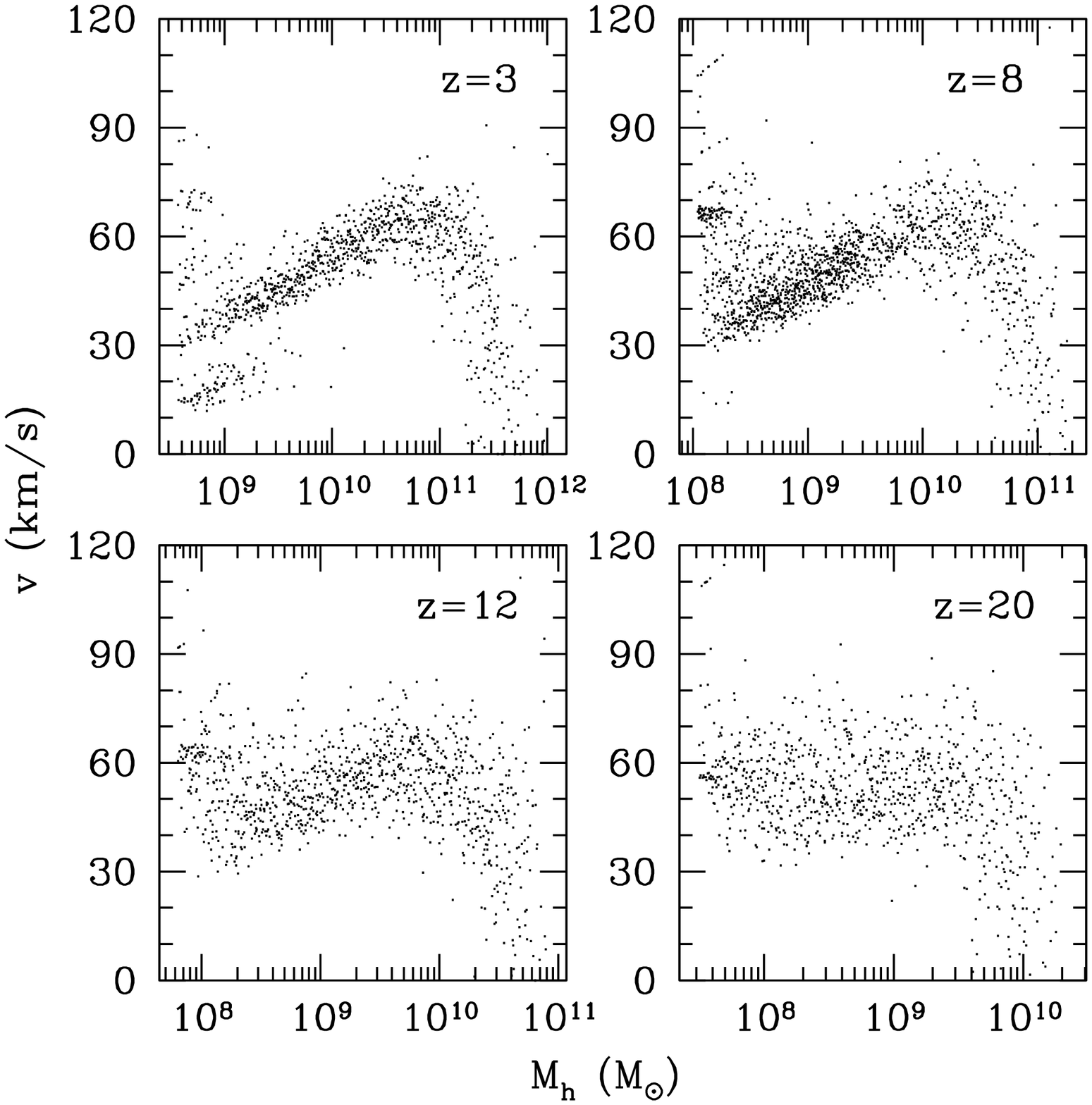}}}}
\figcaption{ Expansion velocity $v_{\rm exp}$ of wind bubbles at various
  redshifts, as a function of halo mass. Note that the mass scale
  changes between the different panels. All panels assume our standard
  set of parameters.
\label{fig:velcomp}}
\vskip 0.2in

\section{Metal Lines at High-Redshift}

We now calculate the distribution of metals within the wind
bubbles, including the most abundant elements and ionization states as
well as the geometry of the metal-polluted regions.  

\subsection{Metal Yields}

The relevant elements are easy to determine: they are those atoms most
abundantly produced by Type II supernovae (for a Scalo IMF) or by pair
instability supernovae (for a VMS IMF).  We summarize the average
yields per supernova $Y_x$ for these elements in Table 1.  For Type II
supernovae, we use the yields of \citet{woosley} averaged over a Scalo
IMF with $Z = 10^{-4} Z_\sun$.  These yields depend on the energy of
the supernova, which introduce uncertainties of $\sim 25\%$.  We have
conservatively taken the yields from the lower energy model.  The
metallicity of the progenitor star introduces a similar uncertainty,
but we do not include chemical evolution within galaxies because other
simplifications in our model create uncertainties at least as large.
For the VMS scenario, we average the yields of \citet{heger} over our
power-law IMF.

Table 1 also lists the properties of several interesting transitions,
including the wavelength $\lambda_m$ and the oscillator strength
$f_{\rm osc}$.  We will focus on these lines in the following.

\subsection{Wind Structure}

Next we must determine how the ions are distributed throughout the
wind.  Note that we assume purely spherical expansion throughout.  In
reality, each halo is surrounded by the ``cosmic web'' of filaments
and voids.  Dense filaments will slow the shell and break spherical
symmetry.  Fortunately, they fill only a small fraction of space and
will not strongly affect the total volume of the bubble (e.g.,
\citealt{theuns}).  A more serious problem is fragmentation (see
below).

Observations of both local starburst winds \citep{hlsa} and LBG
outflows \citep{pettini} show very complex ionization structure, with
absorption from both low ionization states typical of \ion{H}{1}
regions (such as \ion{C}{2}, \ion{O}{1}, \ion{Si}{2}, and \ion{Fe}{2})
and high-ionization states (such as \ion{C}{4} and \ion{Si}{4}).
Within the wind region, the hot gas is inferred to have a very high
covering fraction $C_A$ (defined as the fraction of the
cross-sectional area covered by the gas), while the low-ionization gas
has $C_A \sim 0.4$--$0.8$ depending on the source
\citep{hlsa,shapley}.

If mixing between the ejected gas and the swept-up IGM is efficient,
the thin-shell model requires that a fraction $1-f_m$ of the metals
reside in a cool, dense shell with $T \sim 10^4 \kel$, with the
remainder in the hot, rarefied interior.  Although there is good
evidence for a shell structure in many observations of starburst winds
\citep{martin98,cecil}, other observations suggest that the cool gas
resides in clouds entrained by the wind \citep{hlsa}.  Even though
there is often little evidence for a cool shell in starbursts of the
latter type, the related observations exclusively study young
starbursts in which the wind mass is still dominated by gas ejected
from the host galaxy.  Once the swept-up IGM gas dominates the wind, a
shell will inevitably form (thought it may not be spherical).
It is not clear whether the metals will fully mix with the swept-up
material that constitutes most of the shell mass.  However, in our
model the shell itself travels with $v_{\rm exp} \sim 50 \kms$ while
winds are launched with $v_w \sim 300 \kms$ (see equation
[\ref{eq:vdisk}]).  For $M \ll M_{\rm max}$, entrainment of the
ambient IGM causes most of the deceleration; therefore metals in the
wind will not decelerate significantly until they impact the shell at
$t \sim 5 \times 10^6 (R/{\rm kpc}) \yr$, where mixing through
turbulence or diffusion can occur.

However, the shell may eventually fragment into clouds due to cooling,
Rayleigh-Taylor instabilities, turbulence, or inhomogeneities in the
ambient medium.  Fragmentation is particularly likely given the
relatively modest Mach number of the flow (see \S 2.3).  At $z=3$, an
analysis similar to our own shows that galactic winds must fragment in
order not to overproduce Ly$\alpha$ absorption lines \citep{theuns01}.
If either the shell has fragmented or the metals are distributed in
cold clouds throughout the bubble, only a fraction $C_A < 1$ of all
lines of sight through the wind bubble will produce metal absorption
lines.  On the other hand, each line of sight that does pass through a
cloud will have on average a larger column density of metals.  For
unsaturated lines, the equivalent width $W \propto N_x \propto
C_A^{-1}$, where $N_x$ is the column density of the relevant ion.  In
this regime (which is satisfied for most of the lines of interest; see
\S 4.1), we then have the transformation \bq \frac{d N (>W)}{dz}
\rightarrow C_A \frac{d N (>C_A W)}{d z},
\label{eq:cftransform}
\eq 
where $dN(>W)/dz$ is the number of absorption systems intersected
per unit redshift with equivalent width greater than $W$ along a
random line of sight (see \S 4.3).  For saturated lines, the
equivalent width transformation factor will be smaller.  Because our
results can be scaled relatively easily with $C_A$, for concreteness
we assume that $C_A=1$.  We further discuss the consequences of
fragmentation in \S 5.

The wind structure will also determine the width of the metal lines.
In the shell (with $T \sim 10^4 \kel$), the thermal line widths are
$b_{\rm th} \sim 3.7 (12/A_x)^{1/2} \kms$, where $b$ is the Doppler
parameter and $A_x$ is the atomic mass of the ion $x$.  In this case,
though the lines may be very deep, they will be extremely narrow and
hence difficult to detect except with extremely high-resolution
infrared spectroscopy.  However, low and moderate resolution
spectroscopy of both nearby starbursts and LBG outflows show line
widths comparable to or greater than the inferred mean expansion
velocity of the wind \citep{hlsa,pettini}.  This can be easily
understood if cloud fragments, each with an intrinsic line width
$b_{\rm th}$, are distributed throughout the wind with a velocity
dispersion comparable to $v_{\rm exp}$, or if clouds are continuously
produced and accelerated by the wind.  We assume in the following that
$b=v_{\rm exp}$.  Note that the expansion velocities we find ($b \sim
50 \kms$; see \S 2.3) are much smaller than those typically observed
in young starburst winds.  If star formation is ongoing, and if the
fast wind near the host contains a substantial fraction of the metals,
we will underestimate the true line width.  However, as long as the
lines are not saturated, the equivalent width $W$ is independent of
our choice for $b$.  Even if the lines are saturated, we cannot
overestimate $W$ so long as we underestimate the velocity dispersion.
We emphasize again that the observations from which we calibrate
$b=v_{\rm exp}$ use the starburst galaxy itself as the background
light source; thus, they only constrain the active phases of the
starburst.  The old winds that we study through background quasar
spectra may have very different velocity dispersions.

\subsection{Ionization States}

Given the most abundant elements, we now determine the relevant
ionization states.  Three ionizing processes must be included:
collisional ionization (especially in the hot interior),
photoionization from the host galaxy, and photoionization by the
extragalactic background light.  In general, if a fraction $C_m$ of a given
element is in a specified ionization state, the covering fraction
of the outflow is unaffected but $N_x \propto C_m$.  Thus in the
unsaturated line limit
\bq 
\frac{dN(>W)}{dz} \rightarrow \frac{dN(>C_m^{-1} W)}{dz}.
\label{eq:cmtransform}
\eq
$C_m<1$ can also occur if metals are retained by the host galaxy.
However, simulations suggest that the wind carries a very large
fraction of the metals produced during a starburst \citep{maclow}.

In the thin shell model, we can estimate the ionization state of the
metals in each phase.  We first consider metals in the shell.  If the
hydrogen in the shell is neutral, metals will reside in states typical
of \ion{H}{1} regions (\ion{C}{2}, \ion{O}{1}, \ion{Si}{2}, and
\ion{Fe}{2} for the elements of interest), because photons able to
further ionize these elements will be absorbed by \ion{H}{1}.  The
ionization potential of \ion{O}{1} is $13.62 \ev$, and it remains
locked in charge exchange equilibrium with \ion{H}{1} (see
\citealt{oh} for a discussion of the utility of this mechanism).  The
other elements have first ionization potentials below that of
hydrogen.
We now show that the extragalactic background light most likely
suffices to maintain these first ionized states.  The total column
density of a metal $x$ along a typical line of sight through (each
side of) the wind is 
\bq
N_{x,{\rm tot}} \approx \frac{Y_x}{\omega_{\rm SN}} \,
\frac{\Omega_b}{\Omega_0} 
\, \frac{f_* M_h}{4 \pi A_x m_p R^2}.
\label{eq:nxtot}
\eq 
We can estimate the background flux $J(\nu_{\rm ion}) \equiv
J_{\rm ion}$, where $\nu_{\rm ion}$ is the ionization threshold of the
relevant species, needed to keep a metal ionized by requiring that the
ionization rate exceed the recombination rate within the shell:
$J_{\rm ion} \ga \alpha_x n_e N_{x,{\rm tot}}$, where $\alpha_x$ is
the recombination coefficient and $n_e$ is the electron density in the
shell.  We find
\begin{eqnarray}
J_{\rm ion} & \ga & 5 \times 10^{-31} \left( \frac{n_e}{\bar{n}_b} \right)
\left(
  \frac{f_*}{0.1} \, \frac{126 \msun}{\omega_{\rm SN}} \,
  \frac{\Omega_b/\Omega_m}{0.05/0.3} \frac{M_h}{10^8 \msun} \right)
\nonumber \\
& & \times \left( \frac{Y_x}{0.1 \msun} \, \frac{12}{A_x} \right) \left( 
\frac{10 \kpc}{R} \right)^2 \left( \frac{10}{1+z}  
\right)^3 \left( \frac{0.0245}{\Omega_b h^2} \right), 
\label{eq:jmin}
\end{eqnarray}
in units of $\fluxunits$ and where $\bar{n}_b$ is the mean baryon
density of the IGM.  The electron density is difficult to constrain.
The standard jump conditions for an isothermal shock give a shell
baryon density $n_{\rm sh} \sim {\cal M}^2 \Delta \bar{n}_b$, where
$\Delta$ is the local overdensity through which the shell travels.
However, the shell initially plows through highly overdense material
within the halo ($\Delta \ga 200$) at a very high velocity.  Later,
the shell thickness can only increase at the sound speed and it may
not have had sufficient time to expand to its equilibrium structure.

We now estimate the extragalactic background over the range $\sim
7$--$13 \ev$, which includes photons able to ionize C, Si, and Fe.  We
assume that the emissivity of a typical galaxy obeys $\varepsilon
\propto \nu^{-\alpha}$ and is proportional to the mass of the galaxy.
The volume emissivity of photons is then
\bq
j(\nu,z) = 2 \pi \alpha \hbar N_{\rm ion} f_* n_H^0 \left(
  \frac{\nu}{\nu_H} \right) \left| \frac{dz}{dt} \right| \left|
\frac{d F_{\rm coll}}{dz} \right|,
\label{eq:volem}
\eq 
where $F_{\rm coll}$ is the fraction of collapsed baryons in
galaxies and where we have normalized the emission spectrum to the
number of photons per baryon in stars emitted above the ionization
threshold of hydrogen $\nu_H$, $N_{\rm ion} \sim 4000$
\citep{barkana}.  The extragalactic background is \citep{haiman-h2}
\bq 
J_\nu (z) = \frac{1}{4 \pi} \int j(\nu_z,z') e^{-\tau (z')} d
\ell(z'),
\label{eq:eblform}
\eq 
where $\nu_z = \nu [(1+z')/(1+z)]$, $\tau(z')$ is the IGM optical
depth at $\nu_z$, and $\ell$ is the proper length.  Below $\nu_H$, the
only sources of significant opacity are the Lyman lines of hydrogen.
For a given observed frequency $\nu$ at $z$, the universe is thus
optically thin back to the point $z_{\rm on}$ at which this radiation
would have blueshifted into the nearest Lyman line (denoted here as
Ly$x$), \bq 1 + z_{\rm on} = \frac{\nu_{{\rm Ly}x}}{\nu} (1 + z).
\label{eq:zon}
\eq
Thus (in units of $\fluxunits$)
\begin{eqnarray}
J_{\rm ion}(z) & = & 1.3 \times 10^{-21} \alpha \left(
\frac{f_*}{0.1} \, \frac{N_{\rm ion}}{4000} \, \frac{\Omega_b
  h^2}{0.0245} \right) \left( \frac{\nu_{\rm ion}}{\nu_H}
\right)^{-\alpha}  \nonumber \\
& & \times \int_z^{z_{\rm on}} dz' \left( \frac{1+z}{1+z'}
\right)^{\alpha}  \left| \frac{d F_{\rm coll}}{dz} \right|. 
\label{eq:ebl}
\end{eqnarray}
For carbon, the integral is typically ($\sim 10^{-5},3 \times
10^{-4},2 \times 10^{-3},10^{-2}$) for $z=20,16,12$, and $8$, with weak
dependence on the spectral slope.  The background for Fe and Si is a
few times larger.  Comparing equations (\ref{eq:jmin})
and (\ref{eq:ebl}), we see that so long as the shell is neutral ($n_e
\ll n_{\rm sh}$) \emph{or} the shell is not extraordinarily overdense, the
extragalactic background will easily ionize these elements to
\ion{C}{2}, \ion{Si}{2}, and \ion{Fe}{2}.

By definition, the extragalactic background cannot ionize hydrogen in
the shell until after cosmological reionization.  We now consider
whether the host galaxy can ionize the shell.  According to the
starburst models of \citet{leitherer}, the number of ionizing photons
produced per unit time by an instantaneous starburst with a Salpeter
IMF and metallicity $Z = 0.05 Z_\sun$ (the lowest metallicity they
calculated) a time $t$ after the burst is
\bq
\frac{d N_\gamma}{d t} \sim 10^{51.1} \left( \frac{t}{10^7 \yr} 
\right)^{-4} \left( \frac{f_* M_g}{10^6 \msun} \right) \secinv
\label{eq:sbion}
\eq 
provided that $t \ga 10^7 \yr$.  Here $f_* M_g$ is the stellar mass
produced in the starburst.  Of these, only a fraction $\chi$ can
escape the host galaxy.  This escape fraction is poorly constrained at
present but is likely $\la 10\%$ \citep{leith,wood,dove,steidel}.

As above, we can calculate the column depth of ionized hydrogen
$N_{\rm HII}$ by balancing the number of ionizations and
recombinations per second.  The recombination rate again depends on
the shell density.  $N_{\rm HII}$ will be largest for the smallest
$n_{\rm sh}$; we therefore assume pessimistically that $n_{\rm sh}
\sim {\cal M}^2 \bar{n}_b$.  We compare $N_{\rm HII}$ to the total
column in the shell $N_{\rm sh}$, assuming that the swept up matter
dominates the shell mass.  We then find, for an instantaneous
starburst, that
\begin{eqnarray}
\frac{N_{\rm HII}}{N_{\rm sh}} & \sim 10^{-3} & \left( \frac{f_*}{0.1}
  \, \frac{\chi}{0.1} \, \frac{M_h}{10^8 \msun} \right) \left(
  \frac{20 \kpc}{R} \right)^3 \left( \frac{10}{{\cal M}^2} \right)
  \nonumber \\ \, & \, & \times \left( \frac{10^8 \yr}{t} \right)^{-4} 
  \left( \frac{10}{1+z} \right)^6 \left( \frac{0.05}{\Omega_b} \,
  \frac{0.3}{\Omega_0} \right) \left( \frac{0.7}{h} \right)^4
\label{eq:nhi}
\end{eqnarray}
We have confirmed the approximate validity of this simple estimate
over the range of densities and spectra of interest with CLOUDY
\citep{cloudy}.  Thus, once an appreciable time has elapsed after the
starburst, the shells remain neutral.  However, if active star
formation persists, $d N_\gamma/dt \sim 10^{53.2} (\dot{M}_*/\msun
\yr) \secinv$ \citep{leitherer}.  The shell could then in principle
remain partially ionized for a longer period.  If the shells fragment
or if the metals otherwise reside in clouds, the densities will
increase beyond ${\cal M}^2 \bar{n}_H$, making ionization even more
difficult.  In this case, the ionization balance would also depend on
the size of the clouds, parameterized in our model by $C_A$.

Observations of nearby starburst winds and LBG outflows find strong
absorption from low ionization states \citep{hlsa,pettini}.  In all of
these systems, the starburst is still active and hence $dN_\gamma/dt$
is large.  In the old winds that provide most of the absorption lines,
the arguments above show that low ionization states should be even
more important.  In the following we therefore focus on low ionization
states, and we assume that $C_m=1$.  While this is obviously a naive
approximation, our results can easily be rescaled by equation
(\ref{eq:cmtransform}).  For completeness, we also show some results
for ions appropriate to a ``warm'' medium, \ion{C}{4} and \ion{Si}{4}.
These ions may be appropriate if metals are advected to the shell but
do not mix efficiently with the swept-up IGM, thus remaining at the
interface between the cool shell and the hot cavity.  Lines from these
ions are prominent in LBG winds \citep{pettini,shapley}.

If, however, the metals remain locked in the hot bubble cavity, our
assumptions will break down.  This medium has density $\sim f_m
\bar{n}_H$ and is easily ionized by even trace amounts of star
formation in the host.  More importantly, the gas in this phase has
typical temperatures $T \sim 10^{5.5}$--$10^{6.5} \kel$ and thus is
collisionally ionized to very high ionization states that lack
rest-frame optical or ultraviolet absorption lines.  Fortunately, any
metals carried by the hot wind medium will eventually be advected to
the shell, where cooling and mixing will occur (see \S 3.2).

\section{Results}

With the wind model described in \S 2 and the metal characteristics
outlined in \S 3, we can now calculate the consequences of metal
pollution from galactic winds.  We list the parameters of our model,
along with their standard choices, in Table 2.  Unless otherwise
specified, all curves shown assume these values.  We also consider two
halo cooling channels: atomic hydrogen cooling, which allows halos
with $T_{\rm vir} \ga 10^4 \kel$ to form stars, and molecular hydrogen
cooling, which reduces the threshold to $T_{\rm vir} \ga 400 \kel$.
In the former case, we assume a \citet{scalo} IMF applies, while in
the latter we assume a VMS IMF.  This choice in turn determines
$\omega_{\rm SN}$, $E_{51}$, and $Y_x$.  For H$_2$ cooling, we set
$\omega_{\rm SN} = 462 \msun$ and $E_{51}=10$.

We present the results of our model in the following sections.  First
we describe the observable characteristics of individual winds in \S
4.1.  We then discuss the global evolution of the metal filling factor
in \S 4.2.  Finally, we present our predictions for absorption line
statistics in \S 4.3.

\centerline{{\vbox{\epsfxsize=8cm\epsfbox{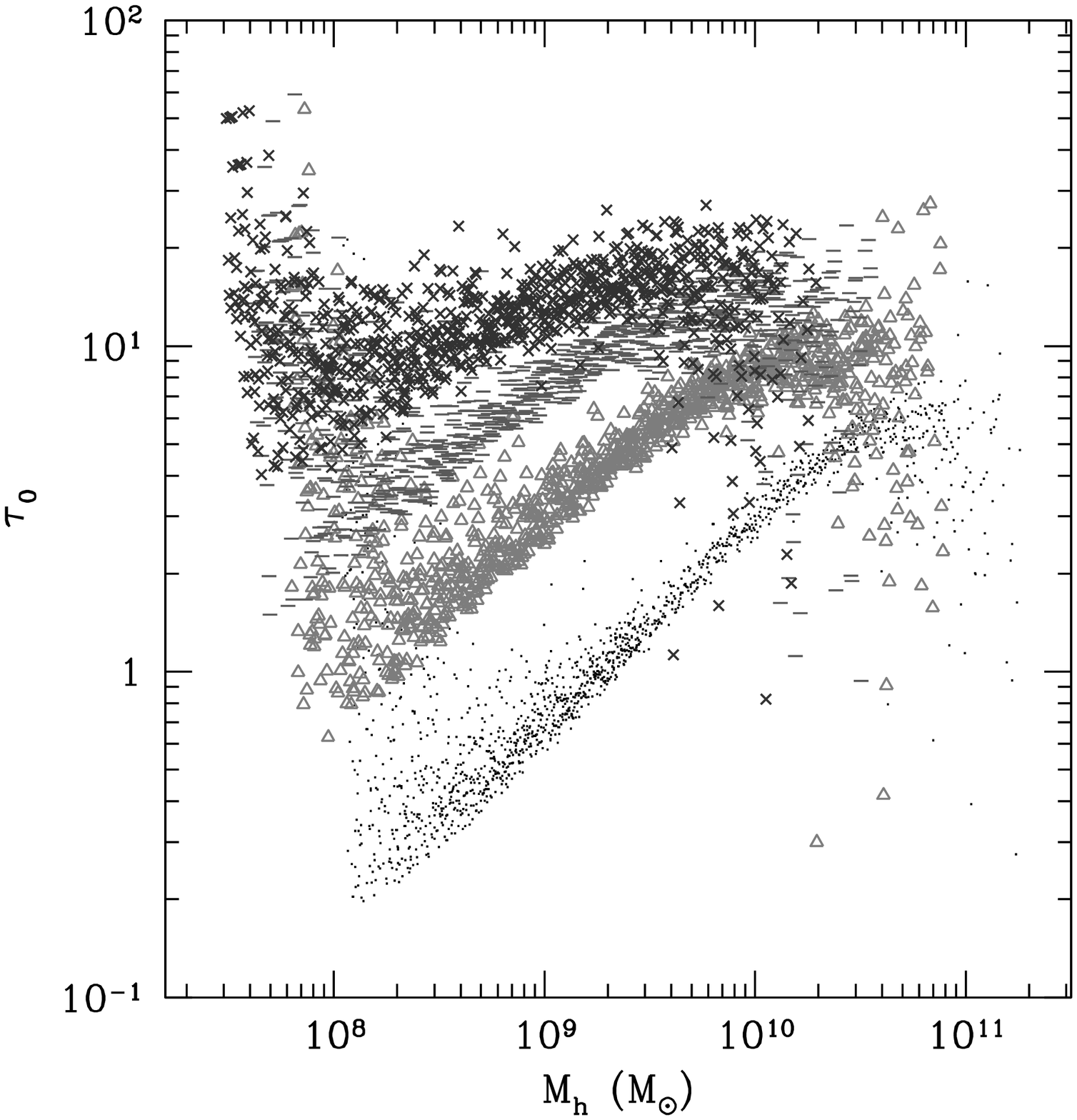}}}}
\figcaption{ Central optical depth $\tau_0$ of the \ion{O}{1} $\lambda
  1302$ feature as a function of halo mass $M_h$ for our standard
  parameter choices.  Crosses, horizontal lines, triangles, and points
  refer to $z=20,16,12$, and $8$, respectively. 
\label{fig:tox}}
\vskip 0.2in

\subsection{Wind Characteristics}

Absorption lines are characterized by three basic parameters (of which
two are independent): the line width $b$, the central optical depth
$\tau_0$, and the equivalent width $W$.  In our model, we set $b =
v_{\rm exp}$ (see Figure 2).  Although we present our results
exclusively in terms of $W$, it is also useful to examine the central
optical depth $\tau_0$ of the absorption systems.  In a real
observation, $\tau_0$ helps to determine the ease with which each line
can be identified.  It also indicates where saturation becomes
important (and hence where equations [\ref{eq:cftransform}] and
[\ref{eq:cmtransform}] break down).  The typical column density of an
ion $x$ through (each side of) the wind per velocity interval is
\bq 
N_x(u) = \frac{N_{x,{\rm tot}}}{\sqrt{\pi} b}
e^{-u^2/b^2},
\label{eq:nxcol}
\eq
where $u$ is the velocity relative to $v_{\rm exp}$.  The optical
depth at line center is then 
\begin{eqnarray}
\tau_0 & = & \frac{\pi e^2 f_{\rm osc} \lambda_m}{m_e c} \,
\frac{N_{x,{\rm tot}}}{\sqrt{\pi} b} 
\label{eq:tau0} \\
& = & 0.2 \left( \frac{f_{\rm osc}}{0.05}
  \, \frac{Y_x}{0.5 \msun} \, \frac{16}{A_x}
  \frac{\lambda_m}{1300 \angstrom} \right)
\left( \frac{50 \kms}{b} \right) \nonumber \\
& & \times  
\left( \frac{f_*}{0.1} \, \frac{126 \msun}{\omega_{\rm SN}} \,
  \frac{\Omega_b/\Omega_0}{0.05/0.3} \, \frac{M_h}{10^9 \msun} \right)
\left( \frac{20 \kpc}{R} \right)^2.
\nonumber
\end{eqnarray}
We show $\tau_0$ as a function of halo mass for several redshifts in
Figure \ref{fig:tox}.  We have chosen the \ion{O}{1} $\lambda 1302$
transition for this plot because it is the strongest low ionization
absorber in our model.  Because $R \propto M_h^{1/5}$ while $b$ is
roughly constant in our model, $\tau_0 \propto M_h^{3/5}$ below
$M_{\rm max}$.  $\tau_0$ increases with $z$ because wind bubbles are
more compact at higher redshifts and hence the typical column density
$N_x$ is much larger.  We see that line saturation is not important
for the low mass-halos that contribute most of the filling factor at
$z \la 12$, but it cannot be neglected at higher redshifts.  At a
given $z$, the minimum $\tau_0$ is determined by the smallest halos
that can collapse to form stars (fixed in this case by the atomic
cooling threshold).

\centerline{{\vbox{\epsfxsize=8cm\epsfbox{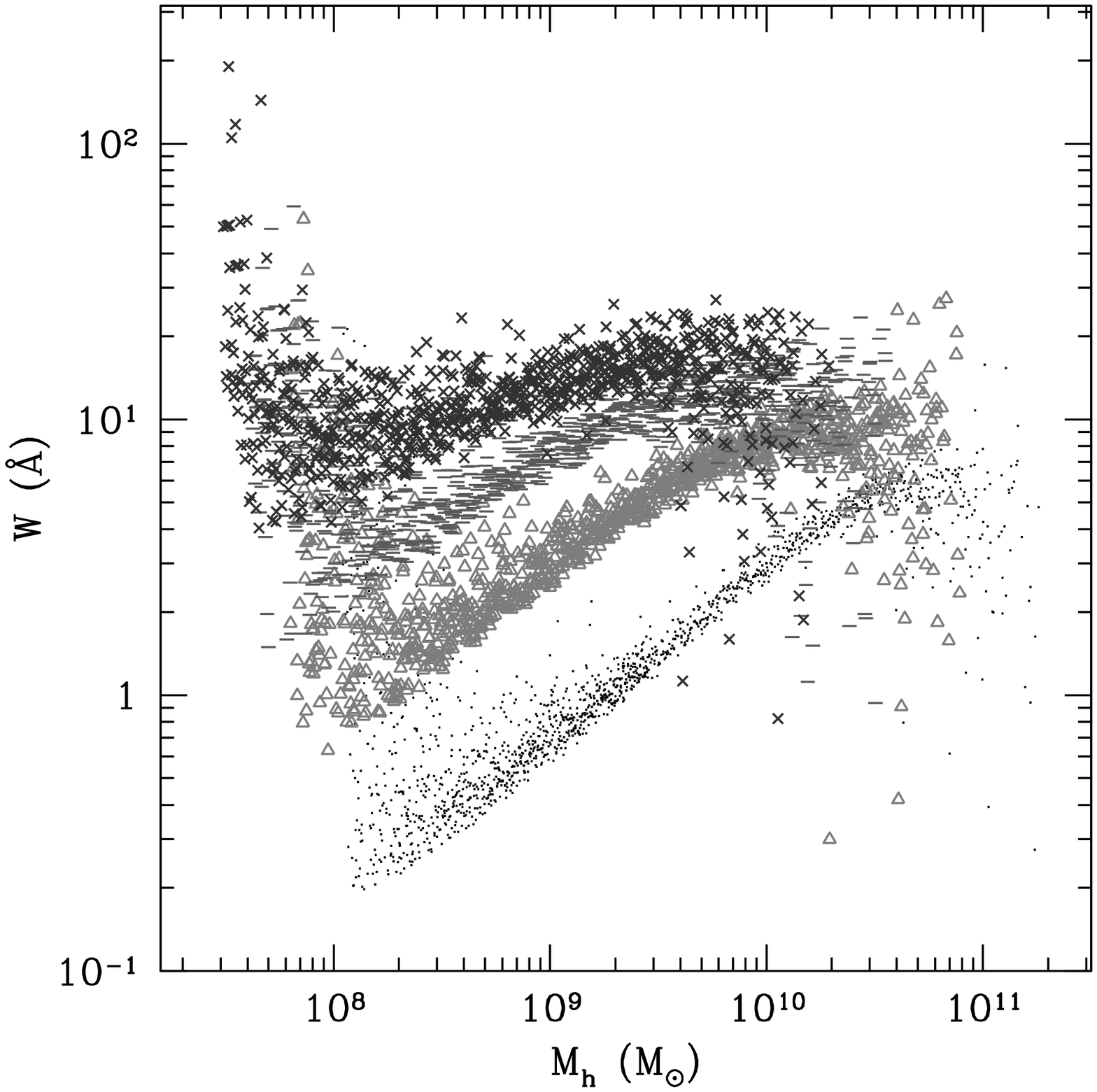}}}}
\figcaption{ Observed equivalent width $W$ of the \ion{O}{1} $\lambda
  1302$ feature as a function of halo mass $M_h$ for our standard
  parameter choices.  Crosses, horizontal lines, triangles, and points
  refer to $z=20,16,12$, and $8$, respectively. 
\label{fig:ewox}}
\vskip 0.2in

The observed equivalent width $W$ is
\bq
W = (1+z) \int \left[ 1 - e^{-\tau(\lambda)} \right] d\lambda. 
\label{eq:eqwid}
\eq
If the line is not saturated, this expression may be written
\begin{eqnarray}
W & \approx & 0.8 \angstrom \left( \frac{f_{\rm osc}}{0.05} 
  \, \frac{Y_x}{0.5 \msun} \, \frac{16}{A_x} \right)
\left( \frac{\lambda_m}{1300 \angstrom} \right)^2
\left( \frac{1+z}{10} \right)
\nonumber \\
\, & \, & \times 
\left( \frac{f_*}{0.1} \, \frac{126 \msun}{\omega_{\rm SN}} \,
  \frac{\Omega_b/\Omega_0}{0.05/0.3} \, \frac{M_h}{10^9 \msun} \right)
\left( \frac{20 \kpc}{R} \right)^2.
\label{eq:widapprox}
\end{eqnarray}
We show $W$ as a function of halo mass in Figure \ref{fig:ewox} (again
for \ion{O}{1} $\lambda 1302$).  Typical equivalent widths increase
from $\sim 0.5$--$10 \angstrom$ between $z=8$ and $z=20$.  $W$
increases with redshift both because of stretching by the cosmological
redshift and because the wind bubbles are smaller at these times.
When line saturation can be neglected, $W \propto b \tau_0 \propto
M_h^{3/5}$.  As the lines become saturated, the relationship flattens.
Interestingly, Figure \ref{fig:ewox} suggests that a line of a given
equivalent width can be associated with a halo in a fairly narrow mass
range, particularly at lower redshifts.  This is essentially a
consequence of the relatively small dispersion in wind bubble sizes
for a fixed set of parameters and our assumption that $f_*$ and
$f_{\rm esc}$ do not vary systematically with halo mass.  As a result,
$N_x$ is well-determined given the halo mass.  There are some recent
indications that $f_* \propto M_h^{2/3}$ \citep{kauffmann,dekel}.  If
so, this would steepen the $M_h$--$W$ relation while preserving the
one-to-one correspondence.

\centerline{{\vbox{\epsfxsize=8cm\epsfbox{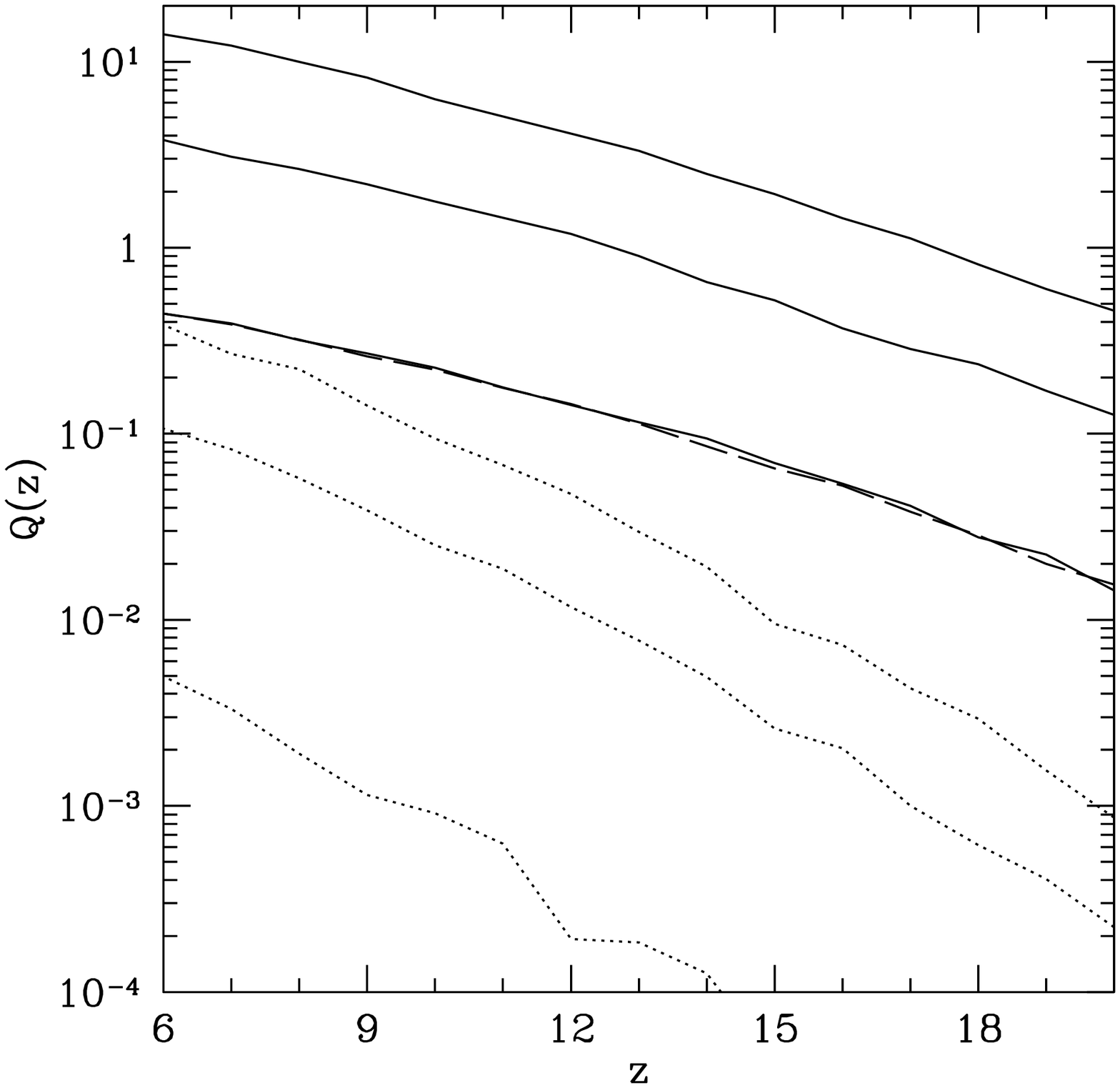}}}}
\figcaption{ Filling factor of polluted regions as a function of
  redshift. Solid curves assume H$_2$ cooling and a VMS IMF, while
  dotted curves assume atomic cooling and a Scalo IMF.  Each set has
  $f_*=0.5,0.1,$ and $0.01$ from top to bottom; other parameters take
  their standard values. The dashed curve assumes H$_2$ cooling with
$f_*=0.1$ and $E_{51}=1$.
\label{fig:ffh2comp}}
\vskip 0.2in

\subsection{Filling Factor}

To calculate the filling factor of the wind bubbles, we compute
\bq
Q(z) = \int_{M_{\rm min}(z)}^\infty dM \frac{dn}{dM} V(M,z),
\label{eq:fillfac}
\eq where $V(M,z)$ is the comoving volume of the wind produced by a
halo of mass $M$ at a redshift $z$ and $dn/dM$ is the halo mass
function \citep{press,sheth}.  Note that we do not include
photoheating due to reionization in determining $M_{\rm min}$.  We
therefore likely overestimate $Q(z)$ for $z < 6$, but we include these
results for comparison with previous work.  Because the star formation
history determines the wind size, we perform the integral using a
Monte Carlo algorithm.  We chose the mass sampling by requiring that
our results converge to $\la 10\%$ between trials.\footnote{For $f_*
  f_{\rm esc} \la 5 \times 10^{-3}$, stochastic fluctuations are large
  and the convergence is somewhat worse; see Figure \ref{fig:ffh2comp}
  for an example.}  Because winds escape small halos most easily,
$Q(z)$ is dominated by winds from low-mass halos (in most of our
calculations, winds with $M_h \la 10 M_{\rm min}(z)$ account for $\sim
75\%$ of the total polluted volume); in practice, the integral need
only extend to some maximum mass $M_{\rm max}$ (see \S 2.2).  The
principal shortcoming of this model is that it neglects the overlap of
wind bubbles; thus, we can have $Q(z) > 1$.  If sources are randomly
distributed, the true filling factor will be $\Phi = 1-\exp[-Q(z)]$.
However, the relationship between $Q(z)$ and the true filling factor
cannot be so easily quantified if sources are clustered (see \S 5).

Figures \ref{fig:ffh2comp} and \ref{fig:ffenergynorm} show $Q(z)$ in
our model.  In Figure \ref{fig:ffh2comp}, we compare $Q(z)$ for H$_2$
cooling (solid curves) and atomic cooling (dotted curves).  Within
each set, the three curves correspond to $f_*=0.01,0.1$, and $0.5$,
from bottom to top.  All other parameters have their standard values.
We see that, in the atomic cooling scenario, $Q(z=12) \approx 1\%$ and
$Q(z=6) \approx 10\%$ (for $f_*=0.1$).  The filling factor increases
rapidly with cosmic time because of the increasing fraction of baryons
in collapsed objects and because the winds have considerably more time
to expand at low redshifts, a point emphasized by
\citet{aguirre-full}.  By allowing less massive halos to form stars
and by injecting an order of magnitude more energy per supernova, the
H$_2$ cooling scenario leads to a much larger filling factor: with
$f_*=0.1$, this model has $Q(z=12) \approx 1$.  The dashed line
(nearly obscured by the lowest solid line) shows the H$_2$ cooling
scenario with $f_*=0.1$ and $E_{51}=1$; thus, even if H$_2$ cooling
creates ``normal'' massive stars the enrichment process is still much
more efficient than with pure atomic cooling.  At very high redshifts,
the minimum halo mass in the atomic cooling scenario is above the
nonlinear mass scale in the \citet{sheth} mass function.  In this
regime, the halo abundance is exponentially suppressed; therefore,
$Q(z)$ for H$_2$ cooling greatly exceeds that for atomic cooling.  The
gap closes (though it is still substantial) as $M_{\rm min}$ falls
below the nonlinear mass scale at lower redshifts.  Our results for
$Q(z)$ in the atomic cooling scenario are very similar to those of
\citet{scann}, even though they use a more sophisticated algorithm to
determine star formation efficiencies and halo clustering.  We find
significant differences, however, between their ``Population III''
scenario and our H$_2$ cooling scenario.  The disparity is likely
caused by their requirement that H$_2$ cooling occurs before disk
formation which imposes a much more stringent mass threshold than
ours, and because we assume that VMS supernovae are more energetic.

\centerline{{\vbox{\epsfxsize=8cm\epsfbox{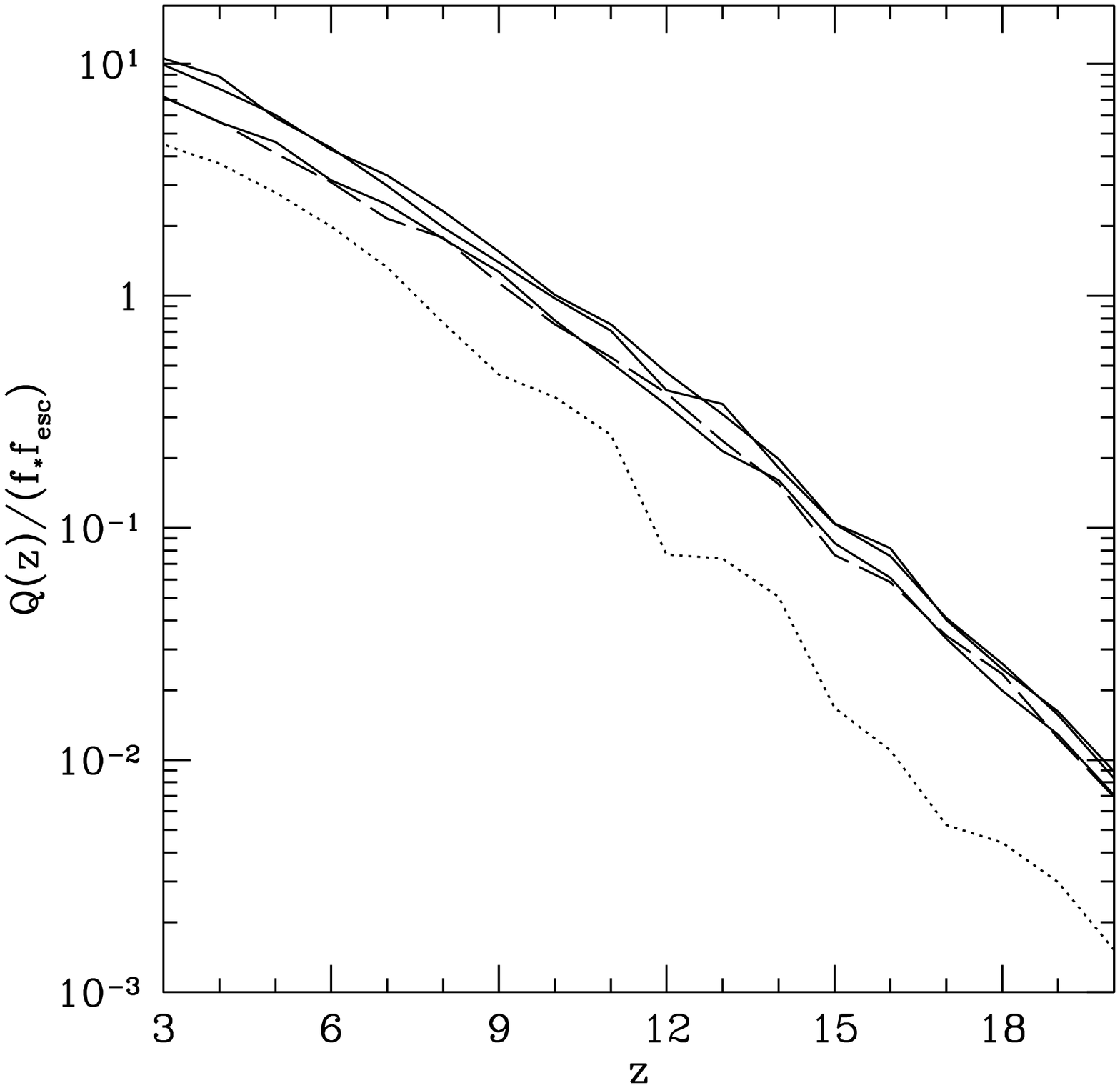}}}}
\figcaption{ Filling factor of enriched regions as a function
  of redshift normalized to $f_* f_{\rm esc}$,
 $Q(z)/(f_* f_{\rm esc})$. The three solid
  curves show $f_{\rm esc}=1,0.25,$ and $0.1$ with $f_*=0.1$.  The
  dashed and dotted curves assume $f_*=0.5$ and $f_*=0.01$,
  respectively, with $f_{\rm esc}=0.25$.  All other parameters take
  their standard values. 
\label{fig:ffenergynorm}}
\vskip 0.2in

Figure \ref{fig:ffenergynorm} plots $Q(z)/(f_* f_{\rm esc})$ for the
atomic cooling scenario.  We have thus normalized $Q(z)$ by the factor
controlling the energy output in the outflows.  This normalized
filling factor is constant (to within a factor $\sim 2$) at a fixed
redshift as long as $f_* f_{\rm esc} \ga 5 \times 10^{-3}$ (c.f.,
\citealt{scann}).  As discussed in \S 2.2, our model yields $R \propto
E^{0.35}$ at a fixed mass $M_h \ll M_{\rm max}$.  Thus, $Q(z)/(f_*
f_{\rm esc})$ should be nearly constant so long as the halos that
contribute most to the filling factor (i.e., the first decade in mass
above $M_{\rm min}$) are well below this mass threshold for efficient
escape.  The proportionality is broken for $f_* f_{\rm esc} \la 5
\times 10^{-3}$ because the number of halos hosting substantial winds
begins to depend on $f_* f_{\rm esc}$.

\subsection{Line Statistics}

Unfortunately, the volume filling factor $Q(z)$ is difficult to
observe, requiring sophisticated pixel-by-pixel analyses (e.g.,
\citealt{aguirre-pixel}).  It is therefore more useful to consider the
statistics of absorption lines.  The number of systems intersected per
redshift interval with an observed equivalent width $W$ greater than
some threshold $W_0$ along a random line of sight is 
\bq
\frac{dN(>W_0)}{dz} = \frac{dr}{dz} \int_{M_{\rm min}(z)}^\infty dM
\frac{dn}{dM} A(M,z) \Theta(W-W_0),
\label{eq:dndz}
\eq
where $A(M,z)$ is the comoving cross-sectional area of the wind,
$dr/dz$ is the comoving distance per unit redshift, and
$\Theta(W-W_0)$ is unity if the equivalent width of a typical line of
sight through the wind is greater than $W_0$ and zero otherwise.
Again, our model neglects halo clustering (see \S 5 for a discussion
of its effects).  We could in principle choose random impact
parameters for each bubble; however, given the approximations we have
already made, we consider this extra layer of complexity to be
unnecessary.  Note that equation (\ref{eq:dndz}) assumes a single
absorption line per halo.  In the ideal case, each intersected wind
bubble would cause 2 lines separated by $\sim 2 v_{\rm exp}$.
However, if the velocity resolution of the observation is lower than
this value, the two lines would appear as a single feature, albeit
with an equivalent width up to twice that which we assume.  The two
lines will also merge if $b > v_{\rm exp}$.  For clarity, we therefore
assume that only one line is produced per halo.

Another important quantity is a measure of the total flux per redshift
interval absorbed by a single transition (neglecting overlap of the
lines),
\bq
W_{\rm tot} = \int W \frac{d}{dW} \left[ \frac{dN(>W)}{dz} \right]
dW. 
\label{eq:wtot}
\eq
If $W_{\rm tot} \ll \lambda_m$, lines will be isolated and easily
separated.  If $W_{\rm tot} \sim \lambda_m$, lines blend together and
absorb a large fraction of the flux from the background source. 

Figure \ref{fig:o1evol} shows the evolution of $dN(>W)/dz$ with
redshift.  Here we show results for \ion{O}{1} $\lambda 1302$, the
strongest line in the atomic cooling scenario.  The curves show line
statistics at $z=20$ (dotted line), $z=16$ (short-dashed line), $z=12$
(long-dashed line), and $z=8$ (solid line).  Each assumes our standard
parameter set.  Clearly, the number of lines intersected depends
strongly on redshift, particularly for weak lines (increasing from
$\sim 0.25$ to $\sim 40$ with $W>0.5 \angstrom$ per redshift interval
from $z=20$ to $z=8$).  Two factors contribute to this increase: the
winds are more compact at higher redshifts (see Figure
\ref{fig:radcomp}) and the fraction of baryons in star-forming
galaxies is much smaller at higher redshifts.  The shape of the curves
are easy to understand.  As shown in Figure \ref{fig:ewox}, a given
$W$ roughly corresponds to halos of a particular mass at fixed $z$.
Thus, at each redshift, $dN(>W)/dz$ is a measure of the (cumulative)
halo mass function weighted by $R^2$.  The ``saturation'' at small $W$
is simply a result of the minimum halo mass for star formation $M_{\rm
min}$.  Figure \ref{fig:o1evol} also shows how $W_{\rm tot}$ evolves
with redshift.  Recalling that \ion{O}{1} is the strongest line we
consider, line blanketing should not be significant for $f_*=0.1$ in
the atomic cooling scenario.

\centerline{{\vbox{\epsfxsize=8cm\epsfbox{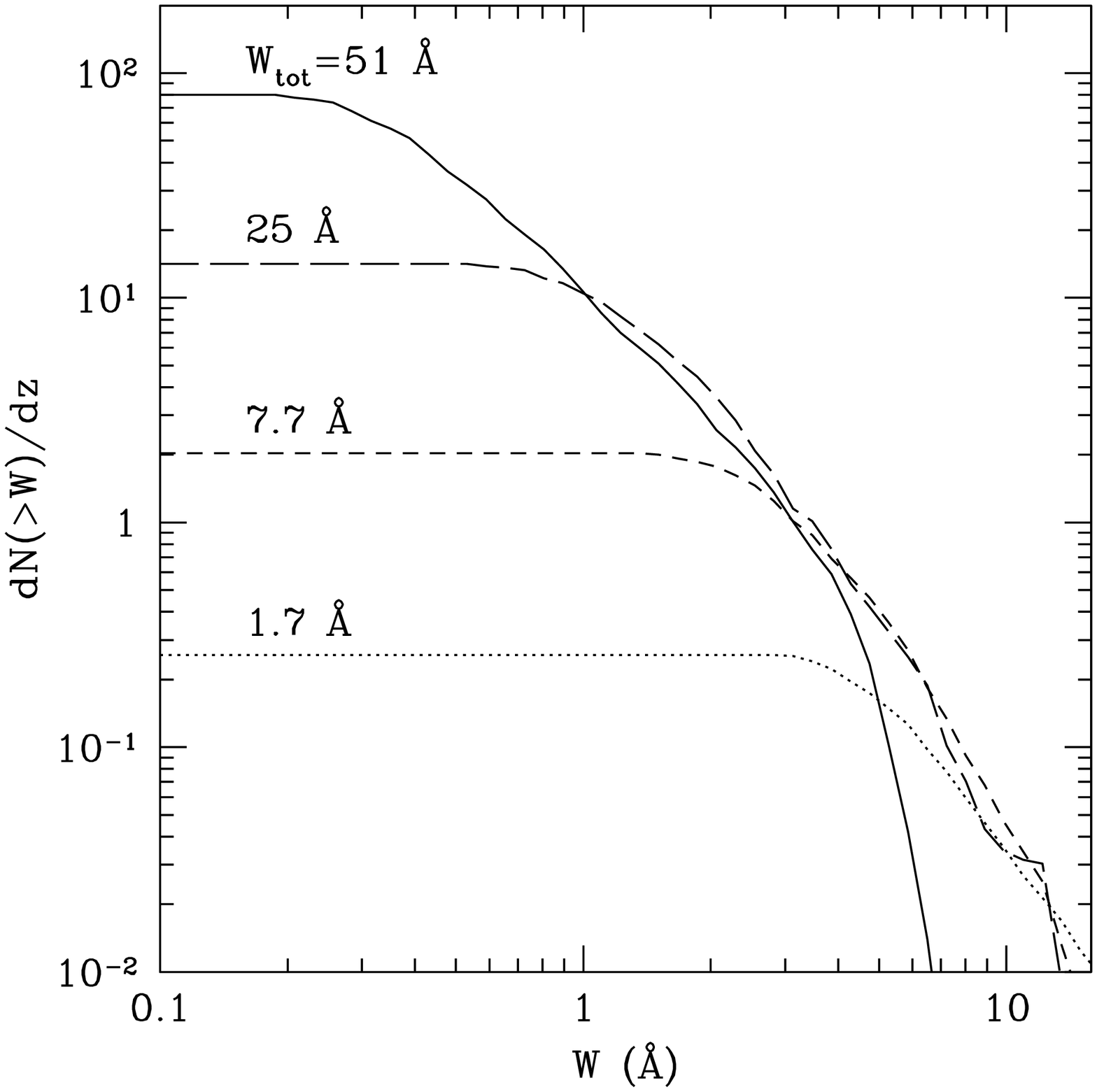}}}}
\figcaption{ Number of intersections per redshift interval above a given
  equivalent width threshold for \ion{O}{1} $\lambda 1302$.  From top
  to bottom, curves refer to $z=8,12,16$, and $20$.  All parameters
  take their standard values.  We also list $W_{\rm tot}$ (see
  equation [\ref{eq:wtot}]) for each line. 
\label{fig:o1evol}}
\vskip 0.2in

Figure \ref{fig:energyz8} shows the \ion{O}{1} statistics for the
atomic cooling scenario at $z=8$.  From top to bottom, the three solid
curves assume $f_*=0.5,0.1,$ and $0.01$, with $f_{\rm esc}=0.25$.  The
two dashed curves assume $f_{\rm esc}=1$ and $0.1$ (top and bottom at
the left, respectively), with $f_*=0.1$.  All other parameters are
standard.  Recall that $Q(z) \propto f_* f_{\rm esc}$; clearly, the
dependence of the line statistics on these parameters is \emph{not}
degenerate.  Essentially, absorption lines give information about both
the extent of the enriched IGM and the level of enrichment.  Recall
that $f_{\rm esc}$ affects only the amount of energy powering the
wind, while $f_*$ affects both the energy input into the wind and the
total mass of metals produced.  Thus, an increase in $f_{\rm esc}$
increases the number of weak lines but \emph{decreases} the number of
strong lines, because the metals become more dilute. An enhanced star
formation efficiency increases the number of lines of all strengths.
The total absorption $W_{\rm tot}$ strikingly reveals the difference.
As shown in Figure \ref{fig:energyz8}, $W_{\rm tot}$ increases rapidly
with $f_*$.  However, it varies only slowly with $f_{\rm esc}$,
increasing from $41 \angstrom$ to $57 \angstrom$ between $f_{\rm
  esc}=0.1$ and $f_{\rm esc}=1$.  Note that if $f_*$ is very high at
this redshift, line blanketing from the many possible transitions
becomes important.

The greatly increased filling factor of polluted regions for H$_2$
cooling compared to atomic cooling (see Figure \ref{fig:ffh2comp})
suggests that these two scenarios may be distinguished through
observations of line statistics.  Figure \ref{fig:ewcool} shows that
this is indeed the case.  We show the statistics for four different
transitions (\ion{O}{1} $\lambda 1302$; long-dashed lines; \ion{C}{2}
$\lambda 1334$, solid lines; \ion{Fe}{2} $\lambda 2344$,\footnote{ We
have selected this line from the several available \ion{Fe}{2}
transitions because it has an ``average'' oscillator strength.  The
strongest line, \ion{Fe}{2} $\lambda 2382$, will have an equivalent
width $\sim 3$ times greater than that shown.} dotted lines;
\ion{Si}{2} $\lambda 1304$, short-dashed lines) at $z=20$ and $z=12$.
The upper set of curves in each panel assumes H$_2$ cooling and a VMS
IMF, while the lower set assumes atomic cooling and a Scalo IMF.  All
other parameters are standard.  We also list $W_{\rm tot}$ for the
\ion{O}{1} $\lambda 1302$ transition in each case.  The difference
between $dN/dz$ for the two models (a factor $\sim 1100$ at $z=20$) is
even more dramatic than the difference in $Q(z)$ (a factor $\sim 500$
at $z=20$).  This occurs because of the rapidly increasing mass
function at small masses, which allows many more individual halos to
produce winds if H$_2$ cooling is permitted; the line statistics are
more sensitive to the number of star-forming halos than is $Q(z)$
because they depend most strongly on the total number of individual
winds that can in principle be intersected.  However, because the
additional halos are small, they generally produce weak lines.
Therefore the enhancement of weak lines is much larger than the
enhancement of strong lines (this being particularly evident at
$z=12$).  Interestingly, $W_{\rm tot}$ is very large for all redshifts
in the H$_2$ cooling scenario.  Thus a large fraction of the
background source flux could be absorbed by overlapping metal lines in
this scenario, extending the apparent Gunn-Peterson absorption to much
higher wavelengths than otherwise expected.

\centerline{{\vbox{\epsfxsize=8cm\epsfbox{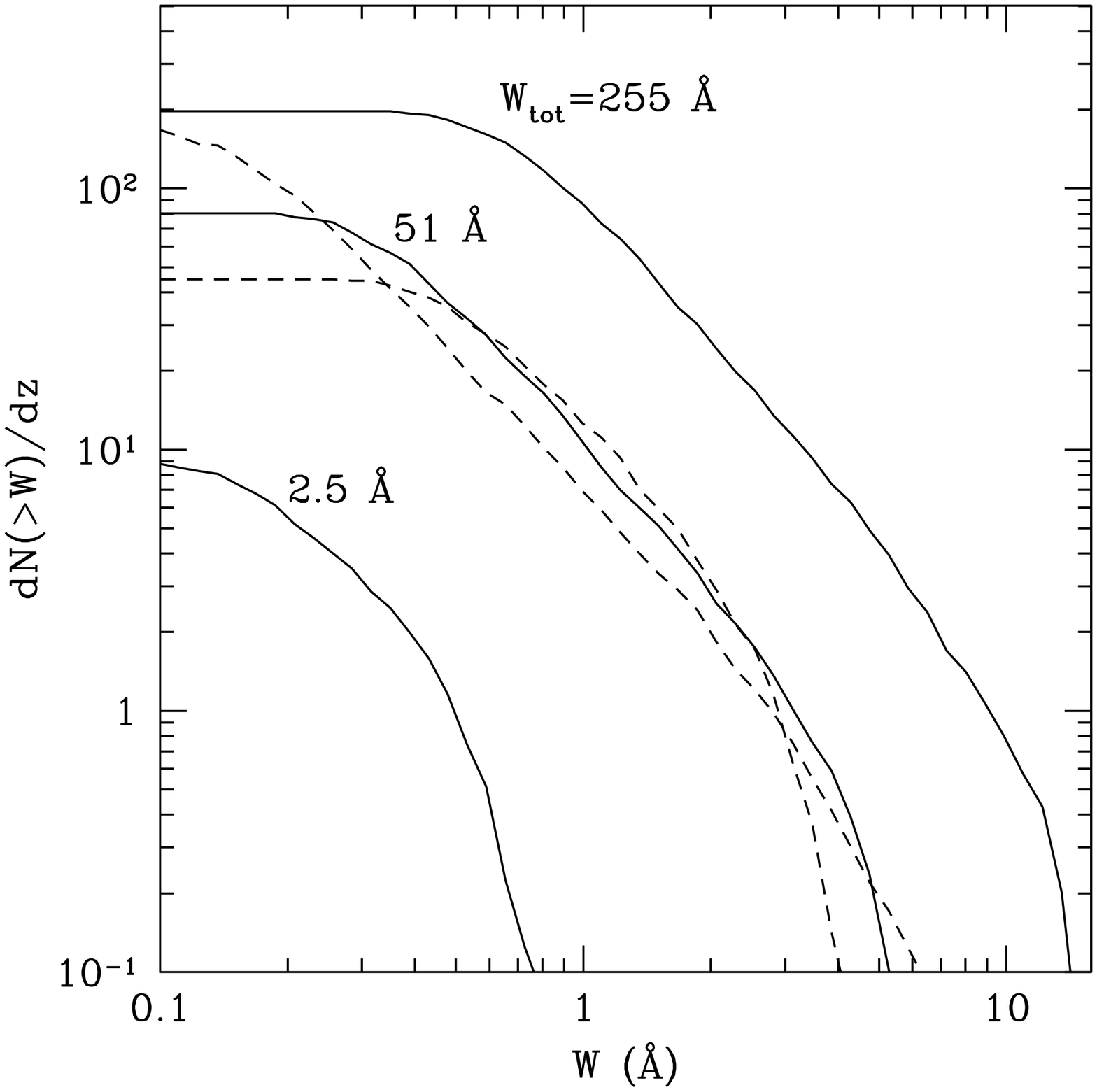}}}}
\figcaption{ Number of intersections per redshift interval above a given
  equivalent width threshold at $z=8$ for the \ion{O}{1} $\lambda
  1302$ transition.  The solid lines assume $f_*=0.5,0.1,0.01$, from
  top to bottom, and $f_{\rm esc}=0.25$.  The dashed lines assume
  $f_{\rm esc}=1$ and $0.1$, from top to bottom at left, and
  $f_*=0.1$. All other parameters are standard. We also list $W_{\rm
    tot}$ (see equation [\ref{eq:wtot}]) for the solid curves.
  $W_{\rm tot}$ varies only slowly with $f_{\rm esc}$ (see text). 
\label{fig:energyz8}}
\vskip 0.2in

We also see the effects of a VMS IMF in Figure \ref{fig:ewcool}.
Although $\omega_{\rm SN}$ varies by less than a factor $\sim 4$
between the two models (and is actually larger for a VMS IMF), the
explosion energy for the VMS IMF is an order of magnitude larger than
for a Scalo IMF.  (Note, however, that even if $E_{51}=1$ for VMS
stars, the differences between the two scenarios are large; see Figure
\ref{fig:ffh2comp}).  The different yields of the two IMFs also cause
differences in the relative line strengths.  For a Scalo IMF, the
strongest features are \ion{O}{1} and \ion{C}{2}, simply because these
are the most abundant elements produced in Type II supernovae.  Pair
instability supernovae, on the other hand, completely disrupt the star
and therefore expel heavy elements much more efficiently
(\citealt{heger}; see Table 1).  As a result, the strongest features
in VMS winds are \ion{O}{1} and \ion{Fe}{2}, with \ion{C}{2} being the
weakest feature.

We argued in \S 3.3 that the low ionization states of metals are
likely to be most important, but it is also possible that many metals
will be in relatively high ionization states.  Figure \ref{fig:ewz8hi}
shows $dN(>W)/dz$ at $z=8$.  The thick curves show low ionization
states (as in Figure \ref{fig:ewcool}) while the thin curves show high
ionization states (\ion{C}{4} $\lambda 1548$, solid line; \ion{Si}{4}
$\lambda 1394$, short-dashed line).  In each case we assume $C_m=1$.
We see that high ionization states actually provide a slightly
stronger signal than the low ionization states; our previous
predictions are therefore conservative in this sense.  However, if the
metals are even more highly ionized, no observable transitions will
remain with $\lambda_m > \lambda_{\rm Ly \alpha}$.

\centerline{{\vbox{\epsfxsize=8cm\epsfbox{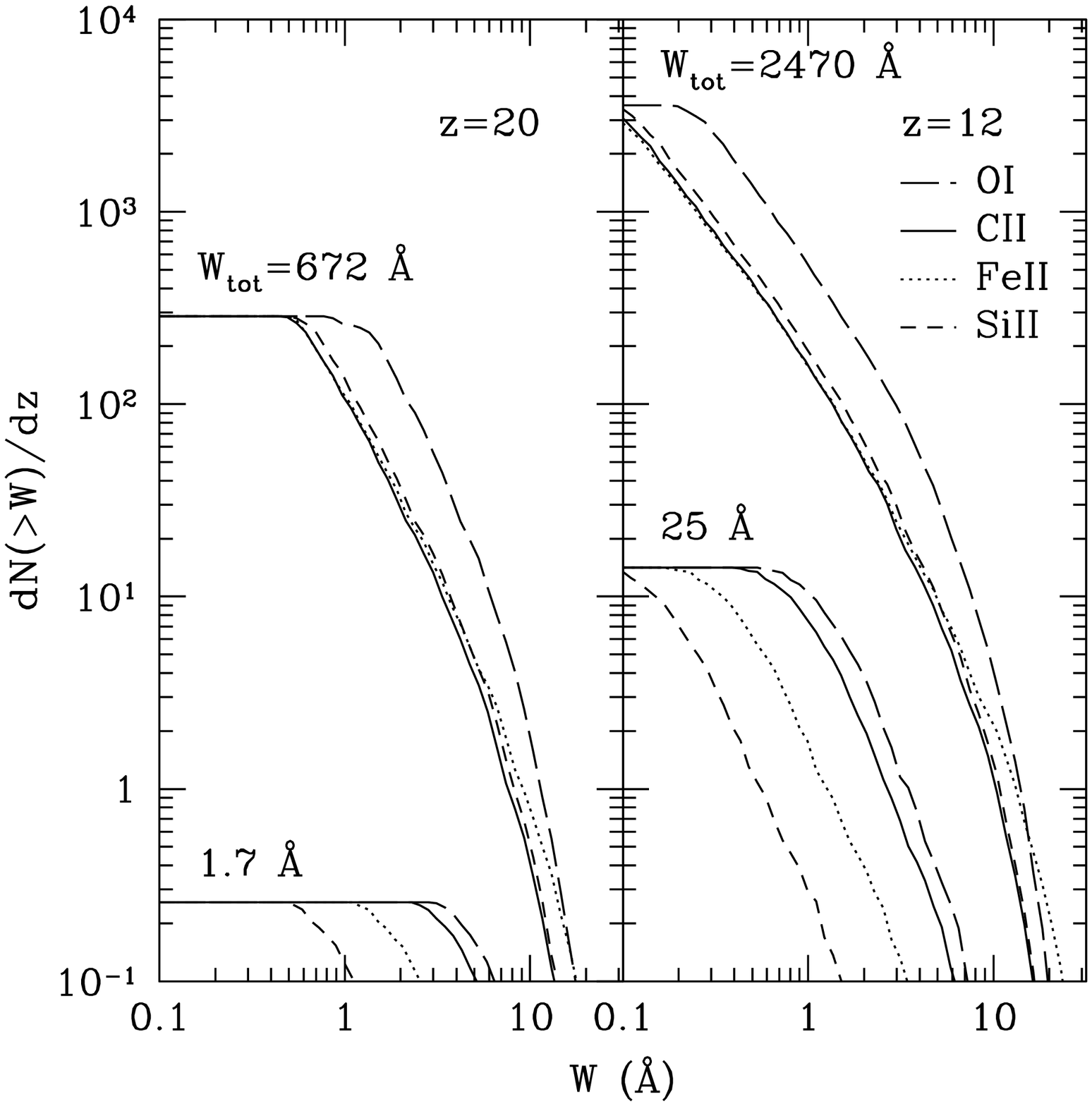}}}}
\figcaption{ Number of intersections per redshift interval above a given
  equivalent width threshold.  In each panel, the bottom set of curves
  assumes atomic cooling while the top set allows H$_2$ cooling.  All
  other parameters take their standard values. We also list $W_{\rm
    tot}$ (see equation [\ref{eq:wtot}]) for the \ion{O}{1} $\lambda
  1302$ transition in each scenario. 
\label{fig:ewcool}}
\vskip 0.2in

Finally, Figure \ref{fig:ewparams} shows the effect of varying other
parameters of the outflow model.  All the curves assume our standard
parameter choices with the following exceptions: $f_d=1$ (dotted
line), $f_{\rm sw}=1$ (dot-dashed line), and no disk formation (dashed
line).  We see that the uncertainty due to model details at $z=8$ is
$\la 30\%$.  The uncertainties in $Q(z)$ are somewhat larger ($\sim
50\%$), because the integrand of equation (\ref{eq:fillfac}) is
proportional to $R^3$, while $dN/dz$ is proportional to the
cross-sectional area $R^2$.  Thus our predictions for line statistics
are fairly insensitive to the detailed wind parameters.  Note,
however, that the uncertainties increase with redshift, reaching a
factor $\sim 2$ by $z=20$ for atomic cooling.  At high redshifts, the
winds have spent a larger fraction of their expansion within the halo
and are hence more sensitive to its structure.  In contrast, at low
redshifts most of the expansion takes place in the IGM and is
independent of assumptions about the halo structure.  Model
assumptions are much less important in the H$_2$ cooling scenario
because the characteristic halo mass (and hence physical scale over
which the halo structure matters) is much smaller.

We have examined the absorption by metals in high-$z$ starburst winds
seen against a background light source.  We allow each star-forming
halo to host a wind, with a time-dependent luminosity determined by
its star formation history.  We use the thin-shell approximation to
trace the size of each wind bubble.  Our model predicts that LBG winds
can reach a proper distance $\sim 100 \kpc$ from the host.  It is
encouraging that this result roughly matches the observationally
inferred distance from LBGs within which \ion{H}{1} is suppressed
\citep{adelberger}.\footnote{Note that we do not include
  photoevaporation or infall suppression after reionization (see
  \citealt{barkana} and references therein) in our model star
  formation histories.  However, LBGs are the most massive objects at
  $z \sim 3$, for which these processes can be ignored.}  Our results
for the filling factor $Q(z)$ are consistent with previous work (e.g.,
\citealt{scann}), but we have emphasized the large uncertainties in
the wind modeling and the correspondingly large uncertainty in $Q(z)$.
We argue that the statistics of metal absorption lines are a useful
way to constrain these parameters.  Not only are these statistics
directly observable, but they are also less dependent on the wind
modeling than $Q(z)$ is.  Our predictions are therefore relatively
robust, although detailed modeling of the wind structure will be
essential for a complete interpretation of the line statistics.

\centerline{{\vbox{\epsfxsize=8cm\epsfbox{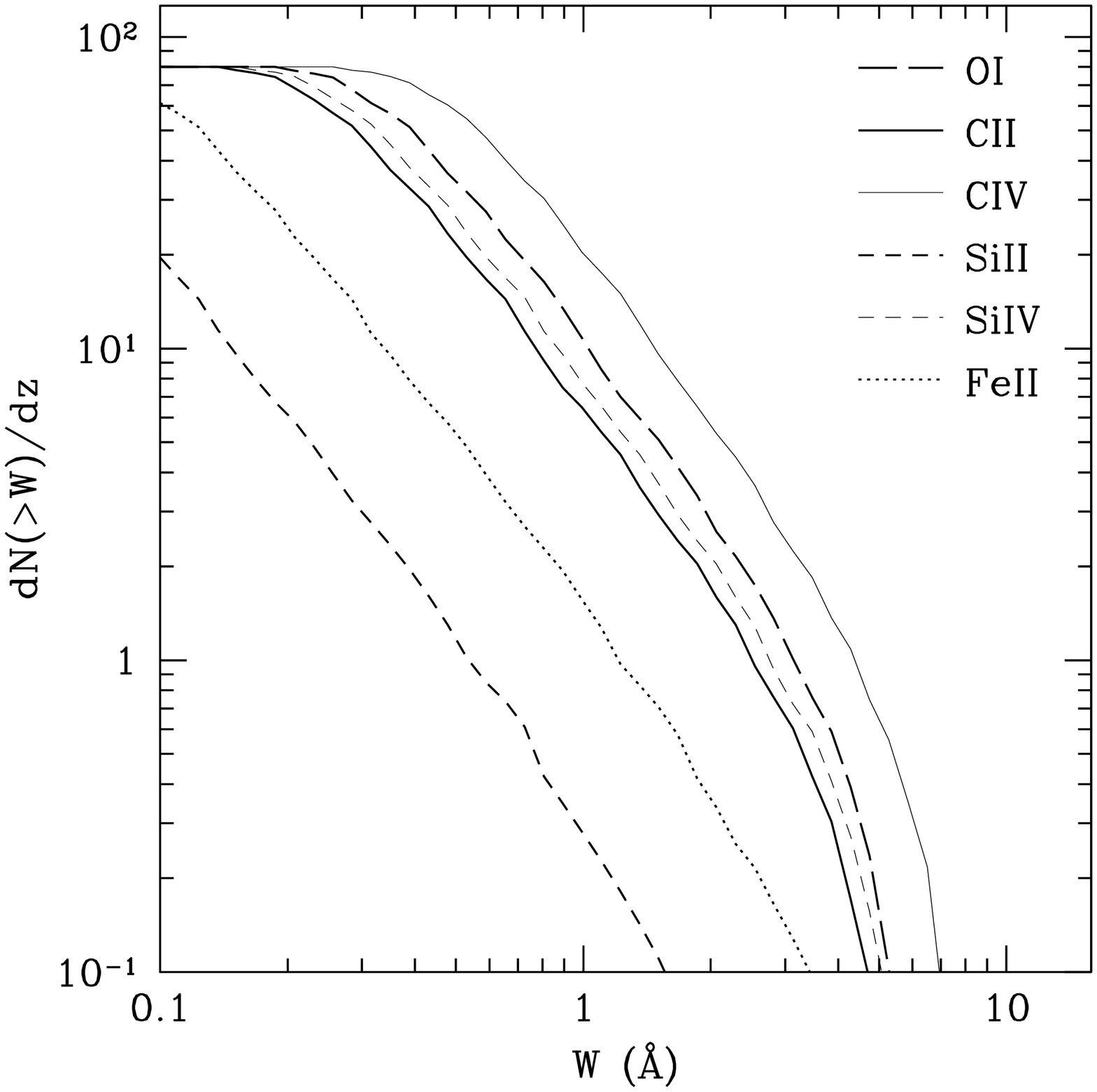}}}}
\figcaption{ Number of intersections per redshift interval above a given
  equivalent width threshold at $z=8$, assuming our standard parameter
  choices.  Thick solid, long-dashed, short-dashed, and dotted curves
  are for \ion{C}{2}, \ion{O}{1}, \ion{Si}{2}, and \ion{Fe}{2},
  respectively.  Thin solid and short-dashed curves are for \ion{C}{4}
  and \ion{Si}{4}, respectively. In each case we fix $C_m=1$. 
\label{fig:ewz8hi}}
\vskip 0.2in

\section{Discussion}

We estimate that $\sim 40$ \ion{O}{1} $\lambda 1302$ lines with
equivalent width $W>0.5 \angstrom$ should be visible per unit redshift
at $z=8$, assuming a star formation efficiency $f_*=0.1$ and
neglecting fragmentation.  This decreases to $\sim 0.25$ lines per
unit redshift by $z=20$, both because of the smaller fraction of
baryons in collapsed objects and because of the brief time that each
wind has had to expand by that time.  The normalization and shape of the
line statistics will primarily constrain $f_*$ and $f_{\rm esc}$, the
fraction of supernova energy that powers the wind.
Because the total amount of absorption is proportional to the amount
of metals ejected, it depends strongly on $f_*$ but is nearly
independent of $f_{\rm esc}$.  The ratio of strong to weak lines, on
the other hand, measures $f_{\rm esc}$ because that parameter controls
the volume over which a fixed metal mass can expand (see Figure
\ref{fig:energyz8}).  

If we allow H$_2$ cooling to occur, then we expect $\sim 250$
\ion{O}{1} $\lambda 1302$ lines per unit redshift at $z=20$, because
this cooling mechanism lowers the threshold halo mass for star
formation and increases the explosion energy per supernova.  In fact,
the combined absorption from the several allowed transitions will
eliminate a substantial fraction of the flux from the background

\centerline{{\vbox{\epsfxsize=8cm\epsfbox{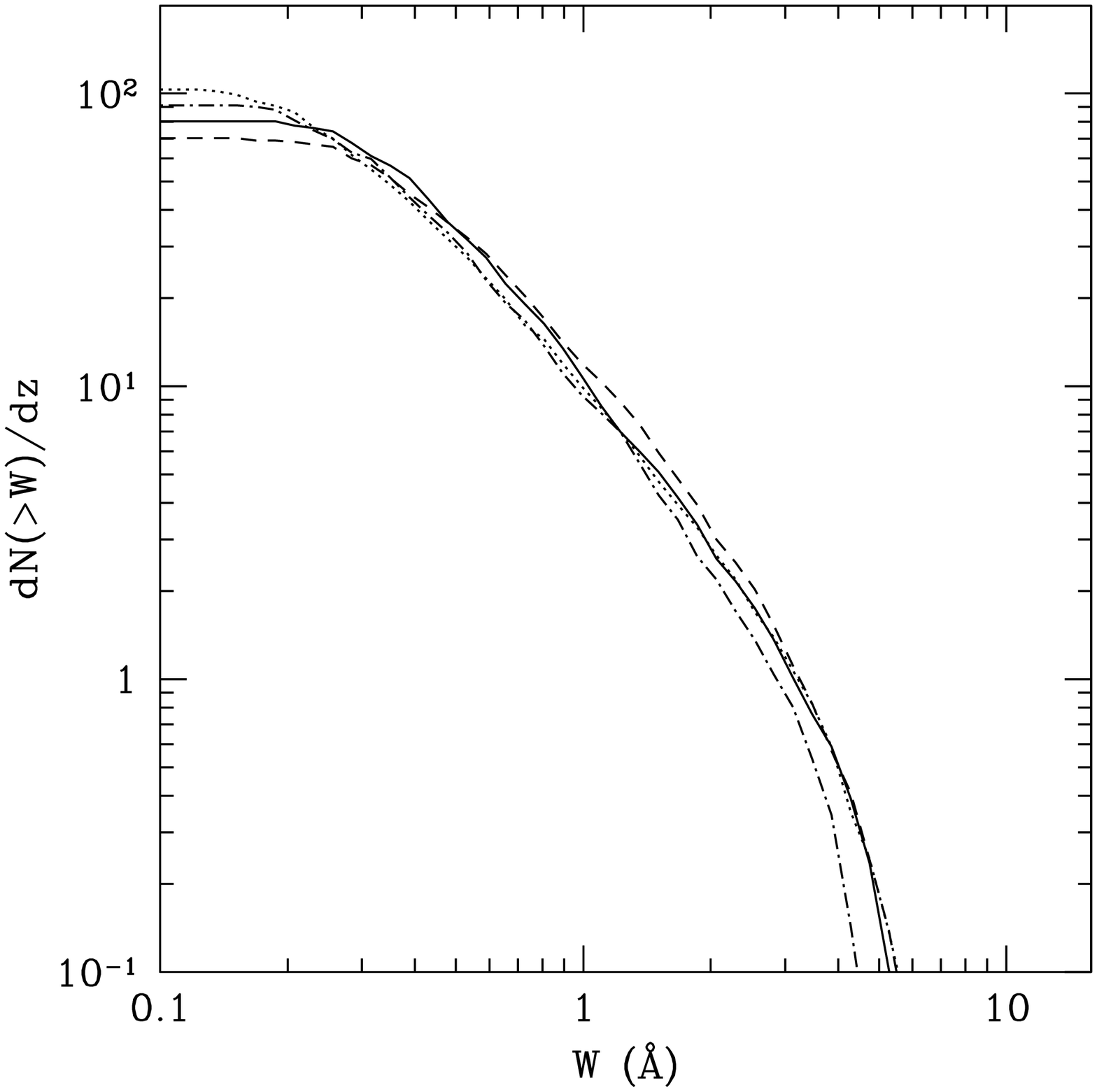}}}}
\figcaption{ Number of intersections per redshift interval above a given
  equivalent width threshold at $z=8$ for the \ion{O}{1} $\lambda
  1302$ transition.  All curves assume our standard parameter set with
  the following exceptions: $f_d=1$ (dotted curve), no disk formation
  (dashed curve), and $f_{\rm sw}=1$ (dot-dashed curve). 
\label{fig:ewparams}}
\vskip 0.2in

\noindent 
source at wavelengths longer than the onset of the Gunn-Peterson
trough.  This would be straightforward to detect, but would also make
isolating individual lines difficult.  The dramatic differences
between this scenario and the atomic cooling scenario are large enough
that the two should be easily separable even given our uncertainties
in $f_*$.  Metal absorption lines therefore offer a possible route to
determine when and if H$_2$ was destroyed throughout the IGM.  This
transition has important implications for the early growth of cosmic
structure (e.g., \citealt{wyithe,mackey}).  The relative strengths of
the observed lines contain information about the IMF.  In particular,
Fe lines will be much more prominent if pair instability supernovae
dominate (i.e., if the IMF is biased toward very massive stars).

Metal absorption line studies can shed light on the chemical evolution
of the universe.  \citet{qian01,qian02} have proposed an important
model for the enrichment history of the universe based on abundance
patterns in Galactic halo stars.  They argue for a ``prompt''
enrichment phase to a threshold iron abundance [Fe/H] $\approx -3$
after which the abundance patterns change markedly; below this
threshold, VMS supernovae dominate the enrichment, whereas
above it ``normal'' supernovae dominate.  The authors implicitly
assume, based on ionic abundances in the Ly$\alpha$ forest
\citep{qsw}, that the prompt phase uniformly enriches all regions of
the IGM to this threshold.  \citet{qsw} find little
evidence for evolution in the elemental abundances in the Ly$\alpha$
forest over the range $z \sim 4.6$--$0.09$, indicating that the prompt
enrichment occurred at $z \ga 5$.  Qualitatively, we have shown that
an early phase of halo formation through H$_2$ cooling forming VMS
stars can pollute large regions of the IGM, lending credence to this
assumption.  Figure \ref{fig:ffh2comp} shows that complete enrichment
of the \emph{entire} IGM by this epoch requires fairly efficient star
formation ($f_* \ga 10\%$) in small halos.  In fact, observations of
the Ly$\alpha$ forest metallicity do not require complete mixing of
metals in the IGM, because these studies are currently only sensitive
to absorbers with $N_{HI} \ga 10^{14.5} \colden$, corresponding to
moderately overdense regions.  Metal enrichment is likely to be
confined in islands that correlate with overdense regions where the
Ly$\alpha$ absorption is enhanced. If VMS regions are preferentially
clustered near cosmic overdensities, it is plausible that those
regions observable as Ly$\alpha$ absorbers with HI column densities
$\ga 10^{14.5} \colden$ are enriched while other regions remain
pristine.  In this case, lower star formation efficiencies would be
allowed.  We would also expect regions polluted by VMS supernovae and
those subsequently polluted by normal supernovae to differ in their
abundance patterns.  An observational probe of the abundance patterns
in the low-density IGM could therefore reveal the era over which the
pollution occurred, although any such measurement would be extremely
difficult.  Note that the \citet{qian01} model assumes a one-to-one,
monotonically increasing relation between metallicity and cosmic time.
If mixing is not complete at high redshifts, the transition from
``very massive'' to ``normal'' star formation may occur at different
times in different regions of the universe.  Thus, the non-uniformity
of the enrichment process, which metal absorption line studies probe
directly, determines the non-uniformity of the transition between
different modes of star formation.

We argued that a substantial fraction of the metals are likely to
exist in low-ionization states (typically \ion{C}{2}, \ion{O}{1},
\ion{Si}{2}, and \ion{Fe}{2}).  If, however, the metals exist
primarily in higher ionization states, such as \ion{C}{4} and
\ion{Si}{4}, then the absorption signal will actually be somewhat
easier to observe because the relevant transitions are stronger (see
Table 1).  Moreover, each of these transitions is a doublet and hence
relatively easy to identify unambiguously.  If the metals
remain in the hot bubble gas with $T \sim 10^6 \kel$, they will reside
in extremely high ionization states (such as \ion{O}{6}) whose
resonant transitions are all blueward of Ly$\alpha$ and therefore
unobservable in the high-$z$ IGM.

The largest uncertainty in constraining $f_*$, $f_{\rm esc}$, H$_2$
cooling, and the IMF will probably be the covering factor of the
enriched material in the wind, $C_A$.  Meaningful constraints on this
factor must await more detailed models of high-$z$ winds, because
inferences from observations of local starburst winds may not apply to
sources in the early universe.  For example, there are indications
that outflows from LBGs are approximately spherical \citep{shapley},
unlike the bipolar winds observed in local starbursts.  Furthermore,
existing observations of absorption by starburst winds and LBG
outflows all use the host galaxy as the background light source and
hence probe only the early phases of the starburst.  Observations
using a background source unrelated to the host, such as a quasar or
GRB, probe winds of all ages, which may have much different structure.
The analysis of \citet{theuns01} implies that winds at $z \sim 3$ must
fragment with $C_A \la 0.01$ in order not to overproduce Ly$\alpha$
forest lines.  It is, however, unclear whether fragmentation of this
degree will occur at the high redshifts we study given the smaller
cosmic times available to the winds.

Unfortunately, uncovering the properties of old winds will be
difficult.  When the host galaxy is used as the background source, the
absorption lines themselves can constrain the structure of the wind
because saturated lines fix the fraction of the galaxy hidden by the
absorbing medium (e.g., \citealt{hlsa}).  However, our background
sources are pointlike and cannot be used in this way.  One strategy to
study the outflow structure is to compare the strengths of the low and
high ionization components of individual winds.  A better strategy is
to compare the absorption along neighboring lines of sight to close
sources on the sky, such as quasar pairs (e.g.,
\citealt{bechtold,crotts}) or multiple images of a single
gravitationally lensed source (e.g., \citealt{rauch}).  Lensed quasar
images are particularly interesting because the lensing probability
increases rapidly with quasar redshift \citep{wyithe-nat,wyithe-lens}.
Such studies can determine the projected extent of individual wind
systems and the distribution of metal clouds within them.

Our model has two important shortcomings.  First, we ignore all
effects arising from large-scale structure.  If star-forming galaxies
are highly clustered (as appears likely at very high redshifts), then
their winds will overlap and decrease the filling fraction of
metal-enriched material.  Clustered sources will also reside in
overdense regions of the IGM, further reducing the wind bubble sizes.
Qualitatively, both these effects will reduce the number of
intersected lines but increase their average strength.  Clustering
will also increase the variance in the number of absorption lines per
unit redshift, because those lines of sight that pass through
overdense regions will have greatly increased numbers of
intersections.  Finally, the filamentary nature of large-scale
structure will affect the geometry of the winds.  Those winds
expanding into filaments will stall earlier, while those expanding
into voids will grow more quickly, resulting in anisotropic growth
\citep{theuns}.  Thus most of the enriched volume will be inside
voids, which would tend to increase the filling fraction.  For
example, with the extreme assumption that all winds expand into a
medium with $\rho_g$ equal to one-tenth the mean density of the
universe, the filling factor of enriched material increases by a
factor $\sim 3$--$5$.  On the other hand, if wind sources are embedded
in overdense regions, the filling factor decreases even more
dramatically because $M_{\rm max}$ decreases rapidly.  Unfortunately,
a detailed treatment of clustering and the anisotropic expansion
around filaments requires considerably more machinery than is
developed here and is best addressed through numerical simulations.
We have chosen a similar wind model to \citet{springel02}, so our
results should help to interpret the effects of the wind model
parameterization in such simulations.

The second shortcoming of our model is the long-term fate of the
winds.  They are likely to fragment due to cooling
instabilities or inhomogeneities in the ambient medium.  If turbulence
is present, the fragments will then mix with the surrounding IGM.
Provided that this mixing occurs at approximately the sound speed, our
estimates of the total filled volume will not be grossly changed;
however, fragmentation could strongly affect the covering fraction of
the metals (see above) and their velocity widths.  

There are at least two types of astronomical objects that can function
as background sources.  The most obvious are quasars.  Bright quasars
have been observed to $z = 6.3$ \citep{fan} and likely extend to $z
\sim 10$ \citep{haiman}.  However, the characteristic quasar
luminosity is expected to decline sharply with increasing redshift
\citep{wyithe-lumfcn} and it would be difficult to find luminous
quasars at much higher redshifts.  Quasars also suffer from
complicated intrinsic absorption, and isolating weak IGM metal lines
may be challenging.  Another possibility is to use gamma-ray bursts
(GRBs) as the background sources.  If GRBs are associated with massive
star formation, some likely occur at extremely high redshifts
\citep{lamb,bromm-grb}.  Cosmic time dilation and a favorable
$K$-correction imply that, at a given time after the burst in the
observer frame, the decrease in brightness with $z$ is mild.
\citet{ciardi} show that the $2 \micron$ flux of a GRB at $z \ga 10$
declines from $\sim 200 \microjy$ 1 hr after the burst to $\sim 1
\microjy$ 10 days after the burst.  GRBs are also attractive sources
because of their intrinsic power-law spectra, from which it will be
straightforward to find absorption lines.  Furthermore, if GRBs occur
in actively star-forming galaxies that drive winds into the
surrounding IGM, the absorption spectrum can probe the host galaxy
wind in detail.  Because GRBs fade, they also serve as useful markers
of faint galaxies at high redshifts.  Subsequent deep imaging of the
host could reveal details that complement the information from
absorption line studies.

Observing IGM metal lines at high redshifts presents a variety of
challenges.  Most importantly, it requires relatively high resolution
spectroscopy of faint sources in the near to mid-infrared ($1 \micron
\la \lambda \la 5 \micron$).  Resolving individual winds requires a
spectral resolution $R \sim c/v_{\rm exp} \sim 6000 \, (50 \kms/v_{\rm
exp})$.  Such resolution is currently achievable with the NIRSPEC
camera on the Keck telescope\footnote{See
http://www2.keck.hawaii.edu:3636/realpublic/inst/nirspec/nirspec.html.}
and with the ISAAC camera on the Very Large Telescope.\footnote{ See
http://www.eso.org/instruments/isaac/.}  However, any ground-based
observation in this wavelength range will suffer from sky
contamination that worsens significantly as the wavelength increases;
thus, ground-based measurements will be limited to $z \la 10$, and
even then will require bright sources.  The James Webb Space
Telescope,\footnote{See http://www.stsci.edu/ngst/.}  due for launch
in the next decade, will have the sensitivity to probe much fainter
background sources with a spectral resolution $R = 5000$.  The
\emph{Mission Simulator}\footnote{See
http://www.ngst.stsci.edu/nms/index.html.}  estimates that a source
with $F_\nu \sim 1 \microjy$ will require a $\sim 16$--$24 \hr$
integration to achieve $S/N=10$ per resolution element at wavelengths
$1.5$--$4 \micron$.  Thus spectroscopy of high-$z$ GRB afterglows with
this instrument will allow us to probe IGM metal lines at $z\ga 10$ in
detail.

Another challenge lies in identifying the lines.  In the case of
\ion{C}{4} and \ion{Si}{4}, the ratio of line strengths in the
doublets will help to confirm the identity of the transition and hence
the redshift of the line.  However, the low ionization states lack
doublet transitions in the relevant wavelength range.  One would
therefore need to identify multiple lines at each redshift in order to
confirm the presence of an individual absorbing wind.  The feasibility
of this technique is then limited not by the strongest lines but by
the weaker transitions.

Finally, we note that absorption line observations such as those we
propose can be combined with imaging surveys to learn even more about
the enrichment process.  \citet{adelberger} have already shown the
usefulness of such an approach for understanding the influence
Lyman-break galaxies have on their surroundings.  Though difficult,
similar observations over a large redshift range have the promise to
map out the history of galaxy feedback on the IGM.

\acknowledgements{

We thank C. Steidel and A. Shapley for illuminating discussions and J.
Schaye for helpful comments that improved the manuscript.  We also
thank the anonymous referee for useful comments.  S.R.F.  thanks the
Institute for Advanced Study, where much of this work was completed,
for kind hospitality. AL acknowledges support from the Institute for
Advanced Study at Princeton, the John Simon Guggenheim Memorial
Fellowship, and NSF grants AST-0071019, AST-0204514.
}

\begin{deluxetable}{cccccc}
  \tablehead{ \colhead{Element} & \colhead{$Y_x$ (Scalo)} 
    & \colhead{$Y_x$ (VMS)} &
    \colhead{Ionization State} & \colhead{$\lambda_m$ (\AA)} &
    \colhead{$f_{\rm osc}$} } 
  \tablecaption{Supernova Yields and Important Transitions} 
  \tablewidth{11.5cm}
  \startdata
  C & 0.1 $\msun$ & 4.1 $\msun$ & \ion{C}{2} & 1334.5 & 0.1278 \\
  & & & \ion{C}{4} & 1548.2\tablenotemark{a} & 0.1908 \\
  & & & \ion{C}{4} & 1550.8\tablenotemark{a} & 0.09522 \\
  O & 0.5 & 44 & \ion{O}{1} & 1302.2 & 0.04887 \\
  Si & 0.06 & 16 & \ion{Si}{2} & 1304.4 & 0.094 \\
  & & & \ion{Si}{4} & 1393.8\tablenotemark{a} & 0.514 \\
  & & & \ion{Si}{4} & 1402.8\tablenotemark{a} & 0.2553 \\
  Fe & 0.07 & 6.4 & \ion{Fe}{2} & 1608.5 & 0.058 \\
  & & & \ion{Fe}{2} & 2344.2 & 0.114 \\
  & & & \ion{Fe}{2} & 2382.8 & 0.300 \\
  \enddata \tablenotetext{a}{Member of doublet}
\end{deluxetable}

\begin{deluxetable}{ccl}
  \tablehead{ \colhead{Symbol} & \colhead{Standard Value} 
    & \colhead{Definition} }
  \tablecaption{Summary of Model Parameters}
  \tablewidth{14cm}
  \startdata
  $f_*$ & $0.1$ & Star formation efficiency \\
  $f_{\rm esc}$ & $0.25$ & Fraction of supernova energy available for
    the wind \\
  $f_{\rm sw}$ & $2 f_*$ & Fraction of host gas entrained in outflow
    \\
  $f_d$ & $0$ & Fraction of swept-up Hubble flow energy available to power
    wind \\
  $f_m$ & $0.1$ & Fraction of swept-up material leaking into bubble
    interior \\
  $\omega_{\rm SN}$ & $126 \msun$ & Mass of stars formed in order to
    produce one supernova \\
  $E_{51}$ & 1 & Explosive energy per supernova, in units of $10^{51}
    \erg$ \\ 
  $Y_x$ & See Table 1 & Yield of element $x$ per supernova \\
  $C_A$ & $1$ & Areal covering fraction of metals \\
  $C_m$ & $1$ & Fraction of metals contained in given ionization state
  \enddata 
\end{deluxetable}

\end{document}